\documentclass[aps,showpacs,prd,twocolumn,nofootinbib,nobibnotes]{revtex4-2}
\usepackage{graphicx}
\usepackage{amssymb}
\usepackage{amsmath}
\usepackage{color}
\usepackage{float}
\usepackage{ulem}
\usepackage{accents}
\usepackage{graphicx}
\usepackage{graphicx}
\usepackage{amsfonts}
\usepackage[colorlinks=true,
pdfstartview=FitV,linkcolor=blue,
citecolor=blue,urlcolor=blue,breaklinks=true]{hyperref}
\usepackage{array}
\usepackage{float}
\usepackage{placeins}
\usepackage[dvipsnames]{xcolor}
\usepackage{csquotes}
\usepackage{bbold}
\usepackage{units}
\usepackage{fixmath}
\usepackage{cancel}

\newcolumntype{C}[1]{>{\centering\arraybackslash}m{#1}}
\renewcommand{\eqref}[1]{\mbox{Eq.~(\ref{#1})}}

\definecolor{ForestForestGreen}{rgb}{0.13,0.55,0.13}
\definecolor{ForestGreen}{rgb}{0.13,0.55,0.13}

\begin{document}

\title{Classical electrodynamics as a tool to examine optical effects of chiral dielectrics and media with magnetic conductivity }
\author{Pedro D. S. Silva$^a$}
\email{pdiego.10@hotmail.com, pedro.dss@ufma.br,}
\author{Manoel M. Ferreira, Jr.$^{a,b}$}
\email{manojr.ufma@gmail.com, manoel.messias@ufma.br}
\affiliation{$^a$Programa de P\'{o}s-gradua\c{c}\~{a}o em F\'{i}sica, Universidade Federal do Maranh\~{a}o, Campus Universit\'{a}rio do Bacanga, S\~{a}o Lu\'is (MA), 65080-805, Brazil}

\affiliation{$^b$Departamento de F\'{\i}sica, Universidade Federal do Maranh\~{a}o,
Campus Universit\'{a}rio do Bacanga, S\~{a}o Lu\'is (MA), 65080-805, Brazil}

\begin{abstract}

We discuss how classical electromagnetic techniques are useful to describe optical effects in conventional and chiral dielectric systems endowed with optical activity. Starting from the Maxwell equations and constitutive relations of the medium, we obtain the wave equation (for the electric field) that yields the refractive indices and the corresponding propagating modes that define the electromagnetic wave polarization. This procedure is employed for bi-isotropic and bi-anisotropic dielectrics, allowing us to determine birefringence, a manifestation of the space or time inversion breaking, typical of the chiral media. Such a method can also be applied for axion electrodynamics and a dielectric supporting an isotropic chiral magnetic current, similar to the Chiral Magnetic Effect (CME). For a diagonal magnetic conductivity, two distinct refractive indices are found, associated with left circularly polarized (LCP) and right circularly polarized (RCP) waves, and yielding circular birefringence (evaluated in terms of the optical rotatory power). Additionally, the scenario of a bi-isotropic medium endowed with isotropic magnetic current is examined, where a rotatory power marked by sign reversal may work as an optical signature of this particular system.

\end{abstract}

\pacs{41.20.Jb, 78.20.Ci, 78.20.Ls, 78.20.Fm}
\keywords{Electromagnetic wave propagation; Optical constants; Magneto-optical effects; Birefringence}
\maketitle

\section{\label{section1}INTRODUCTION}

As well-known, the classical electrodynamics of a continuous medium is stated by the Maxwell equations in matter
\begin{subequations}
	\begin{align}
		\nabla \cdot \mathbf{D} &=\rho\,, \quad \nabla \times \mathbf{H-\partial _{%
				\mathrm{t}}D=J}\,, \label{MINH} \\[1ex]
		\nabla \cdot \mathbf{B} &=0\,, \quad \nabla \times \mathbf{E+\partial _{%
				\mathrm{t}}B=0}\,, \label{MHOM}
	\end{align}
\end{subequations}
where $\mathbf{D}$ and $\mathbf{H}$ are the electric displacement and magnetic field, $\mathbf{E}$ is the electric field and  $\mathbf{B}$ is the magnetic flux density \cite{Jackson,Zangwill,Landau,Fowles,Bain}. Such quantities are related by the constitutive relations (CR), which involve electric and magnetic properties of the medium, such as the electric permittivity $\epsilon$, magnetic permeability $\mu$, and Ohmic conductivity $\sigma$. The displacement vector contains the response
of the matter to applied electric field in the form of the polarization vector, $\mathbf{P}$. Analogously, the field $\mathbf{B}$ includes the magnetization response of the medium, $\mathbf{M}$. Both responses are written in terms of linear relations,
\begin{align}
	\mathbf{P}={\epsilon }_{0}{\chi }_{E}\mathbf{E}, \quad \quad
	\mathbf{M}={\chi}_{M}\mathbf{H},
	\label{PM1A}
\end{align}
where ${\chi }_{E}$ and ${\chi }_{M}$ are the electric and magnetic
susceptibility of the medium, respectively. The relations (\ref{PM1A}) lead to  linear and isotropic constitutive relations, 
\begin{equation}
	\mathbf{D}=\epsilon \mathbf{E}, \quad
	\mathbf{H}=\mu^{-1} \mathbf{B}.  \label{SCR1}
\end{equation}
with
\begin{align}
	\epsilon =\epsilon _{0}(1+\chi^{E}),  \quad \mu =\mu _{0}(1+\chi ^{M}),
\end{align}
being the electric permittivity and the magnetic
permeability of the medium, respectively, obtained from the definitions
\begin{subequations}
	\begin{align}
		\mathbf{D} &={\epsilon }_{0}\mathbf{E}+\mathbf{P}={\epsilon }_{0}(1+{\chi }%
		_{E})\mathbf{E}\,,  \label{eq5B} \\[1ex]
		\mathbf{B} &={\mu }_{0}\mathbf{H}+{\mu }_{0}\mathbf{M}={\mu }_{0}(1+{\chi }%
		_{M})\mathbf{H}.  \label{eq5C}
	\end{align}
\end{subequations}

Using a typical plane wave ansatz, 
\begin{equation}
	(\mathbf{E,D,B,H})=(\mathbf{E}
	_{0},\mathbf{D}
	_{0},\mathbf{B}
	_{0},\mathbf{H}
	_{0})e^{\mathrm{i}(\mathbf{k}\cdot\mathbf{r}-\omega t)},
\end{equation}
the Maxwell equations with $\rho=0$ read
\begin{subequations}
	\label{maxwell-general}
	\begin{align}
		\mathbf{k}\cdot \mathbf{D}& =0\,,\quad \mathbf{k}\times \mathbf{H}+\omega
		\mathbf{D}=-i \mathbf{J}, \\[1ex]
		\mathbf{k}\cdot \mathbf{B}& =0\,,\quad {\bf{k}}\times {\bf{E}}- \omega \mathbf{B} ={\bf{0}}.
	\end{align}
\end{subequations}
Manipulating the Maxwell equations and the CRs, one obtains
\begin{align}
\left[{\bf{k}}\times ({\bf{k}}\times {\bf{E}} ) \right]^{i}-\omega \left({\bf{k}}\times {\bf{B}}\right)^{i} &=0.  \label{extra-Fresnel-1}
\end{align}
The second term of \eqref{extra-Fresnel-1} can be simplified by using the $i$-th  component of Amp\`ere's law,
\begin{align}
\left( {\bf{k}}\times {\bf{B}} \right)^{i}= - \omega \mu \epsilon_{ij}E^{j}, \label{extra-Fresnel-2}
\end{align}
where we have considered a non-conducting medium with the constitutive relations,
\begin{equation}
	 D^{i}=\epsilon_{ij}E^{j}, \,  \,  \, {\bf{H}}= \mu^{-1} {\bf{B}}.
	 \label{DEHB1}
\end{equation}
Then, by replacing \eqref{extra-Fresnel-2} in \eqref{extra-Fresnel-1}, one finds a wave equation for the electric field, given by
	\begin{equation}
		\lbrack \mathbf{k}^{2}{\delta }_{ij}-k_{i}k_{j}-{\omega }^{2}\mu 
					\epsilon_{ij}]E^{j}=0\,,  \label{WE1A}
	\end{equation}
which describes the electromagnetic propagation in the medium. Such an equation provides the refractive indices and the corresponding propagation modes. This method can be applied to conventional and nonconventional dielectrics, including the exotic ones with magnetic conductivity, revealing new propagation properties and associated optical effects.

The Chiral Magnetic Effect (CME) is the macroscopic generation of an electric
current in the presence of a magnetic field as the result of an asymmetry
between the number density of left- and right-handed chiral fermions. It leads to a
current that is linear in the magnetic field~\cite{Kharzeev1}.
This quantum effect has been under extensive research in
particle and field theory as well as nuclear and condensed-matter physics. It was
investigated in quark-gluon plasmas with a chiral chemical potential under
the influence of an external magnetic field \cite{Fukushima,Gabriele} and was
also derived in the context of high-energy physics by Vilenkin (in the
1980's) \cite{Schober,Vilenkin} who supposed an imbalance of fermion chirality
in the presence of cosmic magnetic fields in the early Universe. The CME was
studied in cosmology~\cite{Maxim}, as well, where it was applied to explain
the origin of the very high magnetic field strengths (up to $\unit[10^{15}]{G}$)
observed in neutron stars \cite{Leite,Dvornikov}. An interesting question is the
possible influence of the external axial-vector field $V_{5}^{\mu}$ on the
magnitude of the anomalous current in the CME~\cite{Bubnov}, which was also
examined for the polarization tensor of a photon in a fermion plasma under the
influence of $V_{5}^{\mu}$ \cite{Akamatsu,Maxim1}. 

In condensed-matter systems, the CME has been already observed \cite{Li,Xiaochun-Huang}. It is connected with the physics of Weyl semimetals (WSMs) \cite{Burkov}, where the massless fermions acquire a drift velocity along the magnetic field, whose direction is given by their chirality. Opposite chirality implies opposite velocities, creating a chiral-fermion imbalance that is proportional to the chiral magnetic current. In WSMs, the chiral current may be different from the usual CME linear relation, ${\bf J}={\sigma}_{B}{\bf B}$, when parallel electric and magnetic fields are applied, yielding a current of the type ${\bf J}={\sigma}({\bf E} \cdot {\bf B}){\bf B}$, corresponding to a conductivity effectively proportional to $\textbf{B}^{2}$ \cite{Barnes}. WSMs and CME have been examined in several scenarios and respects, considering the absence of Weyl nodes \cite{Chang}, anisotropic effects stemming from tilted Weyl cones \cite{Wurff}, the CME and anomalous transport in Weyl semimetals \cite{Landsteiner}, quantum oscillations
	arising from the CME \cite{Kaushik}, computation of the electromagnetic fields produced by an electric charge near a topological Weyl semimetal with
	two Weyl nodes \cite{Ruiz}, the chiral superconductivity \cite{Kharzeev2}, wave-packet scattering connected with transport properties \cite{Qing-Dong}, and evaluation of the Goos-H\"anchen lateral shift for reflection at a Weyl semimetal interface \cite{Qing-Dong2}.

The Maxwell-Carroll-Field-Jackiw (MCFJ)
electrodynamics \cite{CFJ,Colladay} has been established in the literature as a pioneering proposal to investigate the possibility of Lorentz violation in vacuum. An interesting connection between the CME and axion electrodynamics with the MCFJ theory has been explored. The CME current can be classically described by the axion Lagrangian \cite{KDeng,Wilczek, Sekine, Tobar, Paixao,Barredo,Qiu},
	\begin{equation}
		\mathcal{L}=-\frac{1}{4}F^{\mu\nu}F_{\mu\nu}+\theta (\mathbf{E}\cdot \mathbf{B)},
	\end{equation}%
	where $\theta$ is the axion field. In this context, the Maxwell equations are
	\begin{align}
		\mathbf{\nabla }\cdot \mathbf{E}&=\rho -\mathbf{\nabla }\theta \cdot \mathbf{B%
		}, \label{M1A} \\
		\mathbf{\nabla }\times \mathbf{B}-\partial _{t}\mathbf{E}&=%
		\mathbf{j}+(\partial _{t}\theta )\mathbf{B}+\mathbf{\nabla }\theta \times 
		\mathbf{E},  \label{M1B}
	\end{align}
	where the terms involving $\theta$ derivatives find association with condensed matter effects \cite{Qiu}. Indeed,  $\mathbf{\nabla }\theta \cdot \mathbf{B}$ represents an anomalous charge density, while $\mathbf{\nabla }\theta \times \mathbf{B}$ appears in the anomalous Hall effect, and  $(\partial _{t}\theta )\mathbf{B}$ plays the role of the chiral magnetic current. For a cold axion dark matter, the associated de Broglie wavelength is large enough to suppose its field does not depend on the space coordinates, $\mathbf{\nabla }\theta
		={\bf{0}}$, so that the Maxwell equations (\ref{M1A}) and (\ref{M1B}) reduce as
	\begin{equation}
		\mathbf{\nabla }\cdot \mathbf{E}=\rho, \quad
		\mathbf{\nabla }\times 
		\mathbf{B}-\partial _{t}\mathbf{E}=\mathbf{j}+(\partial _{t}\theta )\mathbf{B},
		\label{Maxwellaxion1}
	\end{equation}
	where $(\partial _{t}\theta )$ plays the role of the chiral magnetic current. When constant, $(\partial _{t}\theta )=cte$, it can be considered the timelike piece of the MCFJ electrodynamics. The MCFJ Lagrangian is a nice illustration of a Lorentz-violating model that finds applications in condensed matter physics. Lorentz-violating theories have also been applied to describe the Casimir effect in connection with axion electrodynamics \cite{Ruiz1}, Weyl semimetals \cite{Gomez,Ruiz2,Marco}, higher derivative models \cite{Pedro2}, among other interesting cases. 

The MCJF is a type of parity-odd theory that can describe optical active (chiral) media. Chiral matter is usually endowed with space inversion violation \cite{Barron2,Hecht,Wagniere,TangPRL}, being described by  bi-isotropic \cite{Sihvola1,Sihvola2,Sihvola3,Sihvola4} and bi-anisotropic electrodynamics \cite{Kong,Bianiso,Aladadi,Mahmood,Pedro3}, whose constitutive relations read
\begin{subequations}
	\label{constitutive2}
	\begin{align}
		\mathbf{D}& =\hat{\epsilon}\, \mathbf{E}+\hat{\alpha}\,
		\mathbf{B},  \label{constitutive2a} \\
		\mathbf{H}& =\hat{\beta}\, \mathbf{E}+\hat{\zeta}\,\mathbf{B },
		\label{constitutive2b}
	\end{align}
\end{subequations}
and $\hat{\epsilon}=[\epsilon_{ij}]$, $\hat{\alpha}%
=[\alpha_{ij}]$, $\hat{\beta}=[\beta_{ij}]$, and $\hat{\zeta}=[\zeta_{ij}]$
represent, in principle, $3\times 3$ complex matrices. The bi-isotropic relations involve the diagonal isotropic tensors, ${\epsilon_{ij}}=\epsilon\delta_{ij}$, ${\alpha_{ij}}
=\alpha\delta_{ij}$, ${\beta_{ij}}=\beta\delta_{ij}$. In chiral scenarios, LCP and RCP waves travel at distinct phase velocities, implying birefringence and optical rotation \cite{Fowles}. This phenomenon stems from the natural optical activity of the medium or can be induced by the action of external fields (e.~g., Faraday effect \cite{Bennett, Porter, Shibata}), being measured in terms of the rotation angle (per unit length) of the polarization axis, the rotatory power (RP) \cite{Condon}.

This work addresses chiral effects of nonconventional dielectric media by employing usual electrodynamics techniques. We begin by revising basic aspects of electromagnetic propagation in a dielectric medium with absorption, characterized by electric permittivity $\epsilon$, magnetic permeability $\mu$, and Ohmic conductivity $\sigma$. This formalism is also applied to dielectrics described by bi-isotropic and bi-anisotropic constitutive relations, yielding the corresponding refractive indices, propagating modes, and also relevant birefringence effects. Furthermore, the procedure is also implemented to examine the wave propagation in a dielectric modified by magnetic conductivity, which includes usual and bi-isotropic dielectric substrates.

This work is outlined as follows. In Sec.~\ref{section2}, we review basic aspects of electrodynamics in matter composing the classical framework that yields the wave equation, refractive index, and propagating modes. First, we apply the method to discuss media described by bi-isotropic media and bi-anisotropic. In Sec.~\ref{section3}, the electromagnetic propagation in a dielectric with isotropic magnetic conductivity is addressed. To do that, we start from the Maxwell equations in continuous matter,
with the constitutive relations $\mathbf{D}={\epsilon }\mathbf{E}$, $\mathbf{B}={\mu }\mathbf{H}$ and the isotropic chiral magnetic current,
$\mathbf{J}={\Sigma}\mathbf{B}$. We write down the wave equation for the electric field, obtaining the refractive indices and the corresponding polarizations for the propagating modes (which turn out to be RCP and LCP waves). The two real refractive indices for each propagation compose a scenario of circular birefringence, whose non-dispersive rotatory power is carried out. Section \ref{RP_reversion} is dedicated to discuss the electromagnetic propagation in a bi-isotropic dielectric with magnetic conductivity, in which it appears a dispersive RP undergoing sign reversion. Finally, we summarize our results in Sec.~\ref{section6}.

\section{\label{section2}Basic aspects on electrodynamics in continuous matter}

In the following, we revisit the application of the classical electrodynamics formalism to typical systems of matter, such as absorbing, bi-isotropic, and bi-anisotropic dielectrics, to examine how the refractive indices and propagating modes are obtained from the wave equation written from the Maxwell equations and constitutive relations.

\subsection{Usual Ohmic medium}

Considering the Ohmic current, $\mathbf{J}=\sigma\mathbf{E}$, and the relations (\ref{SCR1}),
one obtains 
\begin{equation}  \label{eq10}
	\mathbf{k} \times\mathbf{k} \times\mathbf{E}+{\omega}^{2}\mu \bar{\epsilon}%
	(\omega) \mathbf{E}=0\,,
\end{equation}
where 
\begin{equation}  \label{eq11B}
	\mathbf{k}^{2}={\omega}^{2}{\mu}\bar{\epsilon}(\omega),
\end{equation}
and
\begin{equation} 
	\bar{\epsilon}(\omega)=\epsilon+\mathrm{i}{\frac{{\sigma}}{{\omega}}},
	\label{eq11A}
\end{equation}
is the dispersive and complex permittivity. Equation (\ref{eq10}) implies
\begin{equation}
	\lbrack \mathbf{k}^{2}{\delta }_{ij}-k_{i}k_{j}-{\omega }^{2}\mu {\bar{{%
				\epsilon }}}_{ij}]E^{j}=0\, ,  \label{eq29A}
\end{equation}%
where $\bar{\epsilon}_{ij}=\bar{\epsilon}(\omega)\delta_{ij}$. We then use the definition, 
$\mathbf{k}=\omega \mathbf{n}$, where $\mathbf{n}$ is the vector that yields the refractive index\footnote{Here we
	take into account that the norm $|\mathbf{n}|$ is nonnegative. To permit complex refractive
	indices, we consider $\sqrt{\mathbf{n}^2}$ instead of $|\mathbf{n}|$. The plus sign again
	indicates that we discard refractive indices with negative real parts.}, $n=+\sqrt{\mathbf{n}^2}$, and points along the direction of the wave vector, so that Eq.~(\ref{eq29A}) becomes
\begin{equation}
	M_{ij}E^{j}=\left[n^2{\delta }_{ij}-n^{i}n^{j}-\mu \bar{{\epsilon }}_{ij}%
	\right] E^{j}=0\,. 
	  \label{eq30-0}
\end{equation}
In this case, from Eq. (\ref{eq11B}) we obtain a complex refractive index:
\begin{subequations}
\label{eq13-0}
	\begin{equation}
		\label{eq13}
		\bar{n}=\sqrt{\mu\epsilon +\mathrm{i}{\frac{{\mu\sigma}%
				}{{\omega}}}}=n{^{\prime}}+\mathrm{i}n{^{\prime\prime}}\,,
	\end{equation}
	where
	\begin{equation}
		\label{eq14}
		n^{\prime},n^{\prime\prime}=\sqrt{\sqrt{\left(\frac{\mu\epsilon}{2}\right)^2+\left(\frac{\mu\sigma}{2\omega}\right)^{2}}\pm \left(\frac{\mu\epsilon}{2}\right)}\,\,.
	\end{equation}
\end{subequations}
The imaginary part of the refractive index leads to a real exponential
factor, $e^{-\omega n^{\prime\prime}(\hat{\mathbf{k}}\cdot\mathbf{r})}$
with $\hat{\mathbf{k}}\equiv\mathbf{k}/\sqrt{\mathbf{k}^2}$. This term damps the
amplitude of the wave along the propagation through matter,
which is related to the absorption coefficient ${\alpha}=2{\omega}%
n^{\prime\prime}$, whose inverse value determines the penetration depth.
This is the usual conducting medium scenario.

In general, the dispersion relations and the refractive indices can be obtained from \eqref{eq30-0}, which has non-trivial solutions for the electric field when the determinant of the matrix $M_{ij}$ is null. This condition is basic in the procedure here developed since it provides the dispersion equations needed to analyze the propagation properties of electromagnetic waves in different types of medium. Indeed, the refractive indices of \eqref{eq13-0} are exactly obtained by requiring $\mathrm{det}[M_{ij}]=0$.

Besides the isotropic and linear media described in the simple relations (\ref{SCR1}), there are several distinct types of anisotropic media addressed by general anisotropic CR \cite{Fowles,Zangwill,Landau, Bain}, 
\begin{equation}
	D^{i}=\epsilon _{ij}E^{j}, \quad H^{i}=\mu^{-1}_{ij}B^{j},  \label{SCR1B}
\end{equation}
where $\epsilon _{ij}$ and $\mu _{ij}$ are the permittivity and permeability
tensors, written as 3 x 3 matrices. In this context, it holds
\begin{subequations}
	\label{ep-mu-0}
	\begin{align}
		\epsilon _{ij}& =\epsilon _{0}(\delta _{ij}+\chi _{ij}^{E}),  \label{ep0qui1}
		\\
		\mu _{ij}& =\mu _{0}(\delta _{ij}+\chi _{ij}^{M}),  \label{mu0qui1}
	\end{align}
\end{subequations}
with $\chi _{ij}^{E}$ and $\chi _{ij}^{M}$ representing the susceptibility
	tensors. The expressions in \eqref{ep-mu-0} include the
	polarization and magnetization contributions, 
\begin{equation}
P^{i}=\epsilon _{0}\chi
	_{ij}^{E}E^{j}, \quad M^{i}=\chi _{ij}^{M}H^{j},
\end{equation}
which, inserted in Eqs. (\ref{eq5B}) and (\ref{eq5C}), yields the expressions (\ref{ep0qui1}) and (\ref{mu0qui1}). The simplest permittivity tensor configuration is $%
\epsilon _{ij}=\epsilon \delta_{ij}$, see \eqref{SCR1}, and describes an isotropic medium, like water and glass, where the physical properties do not depend on the direction of the wave propagation. For
anisotropic configurations, the tensor $\epsilon _{ij}$ is nondiagonal and describes uniaxial
and biaxial crystals \cite{Fowles,Landau, Bain}, which display
optical activity (chirality) \cite{Hecht,Wagniere} and birefringence \cite{Condon}.

\subsection{Bi-isotropic medium}

Some materials present additional electric and magnetic responses, encoded in extended linear constitutive relations
	\begin{subequations}
		\label{constitutive-relations-biisotropic-1}
\begin{align}
\mathbf{D} ={\epsilon}\mathbf{E}+{\alpha}\mathbf{B}, \label{CRBI1A} \\
 \mathbf{H} ={\beta}\mathbf{E}+\frac{1}{\mu }\mathbf{B}, \label{CRBI1B}
\end{align}
	\end{subequations}
including $\alpha $, $\beta$ as complex magnetoelectric parameters that measure the response of the medium to the applied fields: while $\alpha$ measures the electric response to the
magnetic field, $\beta$ represents the magnetic response to the electric field. In order to be consistent with energy conservation \cite{Pedro3}, it must hold 
\begin{equation}
	\alpha=-\beta^{*}.  \label{SCR2}
\end{equation} 
The relations (\ref{constitutive-relations-biisotropic-1}) describe the electromagnetic response of bi-isotropic media (the most general linear,
homogenous, and isotropic materials). The bi-isotropic relations (\ref{CRBI1A}) and (\ref{CRBI1B}) have been investigated both theoretically \cite{Sihvola1,Sihvola2,Sihvola3, Kong} and experimentally \cite{Rado, Aladadi, Jelinek}, being also important to address optical properties of topological insulators \cite{Zou,Urrutia, Lakhtakia, Winder,Li1,Tokura,Li-Cao}. Bi-isotropic relations are relevant for axion electrodynamics \cite{Sekine, Tobar, BorgesAxion}, construction of optical isolators from chiral materials \cite{Silveirinha}, Casimir effect in chiral media \cite{Casimir,Casimir2,Casimir3}, and surface plasmon polaritons \cite{Darinskii}.  Magneto-optical effects are used to investigate features of new materials \cite{Chang1, Tse} and graphene compounds \cite{Crasee}.

The constitutive relations (\ref{CRBI1A}) and (\ref{CRBI1B}), when inserted  in the Maxwell equations, provides
\begin{align}
	\left[ \mathbf{k}\times \left(\mathbf{k}\times \mathbf{E}\right)\right]
	^{i}+\omega ^{2}\mu \bar{\epsilon}_{ij}E^{j}=0,  \label{ex8} \\
	\lbrack \mathbf{k}^{2}{\delta }_{ij}-k_{i}k_{j}-{\omega }^{2}\mu {\bar{{%
				\epsilon }}}_{ij}]E^{j}=0\,, \label{eq29B}
\end{align}
with the effective permittivity tensor,
	\begin{equation}
	\bar{\epsilon}_{ij}= \epsilon \delta _{ij}+(\alpha +\beta ) \varepsilon_{ijm}  n_{m},
	\label{iso4}
	\end{equation}
	where the last term in RHS represents the ``magnetic-electric" response of
	the medium and $\epsilon_{ijk}$ is the Levi-Cevita symbol with $\epsilon_{123}=1$. From \eqref{ex8} one obtains $k_i\bar{\epsilon}_{ij}E^{j} =0$, so that
	 $\bar{D}^i=\bar{\epsilon}_{ij}E^j$ can read as an extended and transversal displacement vector, fulfilling  $\mathbf{k} \cdot \bar{\mathbf{D}} =0$.  Here, the electric displacement vector can be written as
	 \begin{align}
	 	\bar{{\bf{D}}} &= \epsilon {\bf{E}} + \frac{\alpha}{\omega} {\bf{k}}\times {\bf{E}}.\label{DDiso}
	 \end{align}
The Gauss' law, ${\bf{k}}\cdot {\bf{D}} =0$, implies transversal modes, that is, ${\bf{k}}\cdot {\bf{E}}=0$.

	As we have started from a bi-isotropic CR, in which $\epsilon_{ij}=\epsilon \delta
	_{ij} $, $\mu_{ij}^{-1}=\mu ^{-1}\delta _{ij}$, $\alpha_{ij}=\alpha \delta _{ij}$, $\beta_{ij}=\beta \delta _{ij}$, any anisotropy effect comes from the extended structure of the
	constitutive relations of \eqref{constitutive-relations-biisotropic-1}.
	Equation (\ref{ex8}) can also be written as
	\begin{equation}
		M_{ij}E^{j}=0,  \label{ex11}
	\end{equation}
with the tensor $M_{ij}$ given as
	\begin{equation}
		M_{ij}=n^{2}{\delta }_{ij}-n_{i}n_{j}-\mu \bar{{\epsilon }}_{ij},
		\label{ex12}
	\end{equation}
In this case, the tensor $M_{ij}$ (\ref{ex12}) has the form
	\begin{align}
	M  \equiv [M_{ij}] &= \mathcal{N}  - \mu (\alpha+\beta) \begin{pmatrix}
		0 & n_{3} & - n_{2} \\
		- n_{3} & 0 & n_{1} \\
		n_{2} & - n_{1} & 0
	\end{pmatrix} ,  \label{eq62A}
	\end{align}
	where
	\begin{equation}
	\mathcal{N}=\begin{pmatrix}
		n_{2}^{2}+n_{3}^{2}-\mu\epsilon  & -n_{1}n_{2} & -n_{1}n_{3} \\
		-n_{1}n_{2} & n_{1}^{2}+n_{3}^{2}-\mu\epsilon  & -n_{2}n_{3} \\
		-n_{1}n_{3} & -n_{2}n_{3} & n_{1}^{2}+n_{2}^{2} -\mu\epsilon
	\end{pmatrix} , \label{Nij1}
	\end{equation}	
represents the contribution of a usual dielectric. Requiring $\mathrm{det}[M_{ij}]=0$, one gets the nontrivial solution, 
	\begin{equation}
	n^{4}-n^{2}\left[ 2\mu\epsilon  -\mu ^{2}(\alpha +\beta )^{2}\right] +\mu ^{2}\epsilon^{2}=0.  \label{iso6}
	\end{equation}%
yielding the following refractive indices:
	\begin{align}
	n_{\pm }^{2}=& \mu{\epsilon}- 2Z \mp \mathrm{i}\mu (\alpha +\beta )%
	\sqrt{\mu\epsilon -Z},  \label{iso7}
	\end{align}
	where
	\begin{equation}
	Z=\frac{\mu ^{2}(\alpha +\beta )^{2}}{4}.  \label{iso-7-1}
	\end{equation}
	Thereby the corresponding $n_{\pm }$ are
	\begin{equation}
	n_{\pm }=\sqrt{\mu\epsilon -Z} \mp \mathrm{i}\sqrt{Z},
	\label{iso10A}
	\end{equation}
	where we have considered only the indices with a positive real piece in order
	to avoid metamaterial behavior. As shown in \eqref{iso10A}, each refractive index $n_{+}$ or $n_{-}$ is the same for any propagation direction since the bi-isotropic system does not have a preferred direction. Despite that, the system may manifest circular birefringence due to the way the fields are coupled. In the
	limit $(\alpha +\beta )\rightarrow 0$, one recovers the refractive index
	of an isotropic dielectric medium, given by
	\begin{equation}
		n_{\pm }^{2}=\mu \epsilon .  \label{iso8}
	\end{equation}
	
	\subsubsection{Propagation modes} 
	
The propagating modes stem directly from \eqref{ex11}. Let us choose
a convenient coordinate system for a wave propagating in the z-axis, 
\begin{equation}
	\mathbf{n}=(0,0,n_{3}), \label{iso-prop-1A}
\end{equation}
Replacing the simple choice (\ref{iso-prop-1A}) in the matrix (\ref{eq62A}),
	\begin{equation}
		M=%
		\begin{pmatrix}
			n_{3}^{2}-\mu \epsilon &  & -\mu (\alpha +\beta )n_{3} &  & 0 \\
			\mu (\alpha +\beta )n_{3} &  & n_{3}^{2}-\mu \epsilon &  & 0 \\
			0 &  & 0 &  & -\mu \epsilon%
		\end{pmatrix}%
		,  \label{matrix-isotropic-modes-1}
	\end{equation}%
	and implementing the refractive indices (\ref{iso10A}), the condition $
	M_{ij}E^{j}=0$ provides the following normalized solutions of the electric
	field of the propagating modes:
	\begin{equation}
		\hat{\mathbf{E}}_{\pm }=\frac{1}{\sqrt{2}}%
		\begin{pmatrix}
			1 \\
			\mp \mathrm{i} \\
			0%
		\end{pmatrix}%
		,  \label{iso-prop-2}
	\end{equation}%
	where
	$\hat{\bf{E}}_{+}$ and $\hat{\bf{E}}_{-}$ represent
	the right-handed circular polarization (RCP) and left-handed circular polarization (LCP) vectors, respectively\footnote{We define a polarization as right-handed (left-handed) if the
polarization vector of a plane wave rotates along a circle in clockwise (counterclockwise)
direction when the observer is facing into the incoming wave \cite{Jackson, Zangwill}.}. The solution (\ref{iso-prop-2}) does
	not depend on the nature (real or complex) of the parameters $\alpha$ and
	$\beta$, in such a way it will be valid for all the cases examined in this
	section. 
	
	We point out that the circular polarization solution (\ref{iso-prop-2}) is not exclusive of the $z$-propagation direction. Indeed, taking on the propagation in the $x$-axis, $\mathbf{n}=(n_{1},0,0)$, the matrix (\ref{eq62}) takes the form,
	\begin{equation}
		M=%
		\begin{pmatrix}
			-\mu \epsilon &  & 0  &  & 0 \\
			0 &  & n_{1}^{2}-\mu \epsilon &  & -\mu (\alpha +\beta )n_{1} \\
			0 &  & \mu (\alpha +\beta )n_{1} &  & n_{1}^{2}-\mu \epsilon%
		\end{pmatrix},  \label{matrix-isotropic-modes-1B}
	\end{equation}
	whose associated modes,
	\begin{equation}
		\hat{\mathbf{E}}_{\pm }=\frac{1}{\sqrt{2}}%
		\begin{pmatrix}
			0 \\
			\pm \mathrm{i} \\
			1%
		\end{pmatrix}%
		,  \label{iso-prop-2B}
	\end{equation}
	also correspond to transversal circularly polarized waves.
	
	\subsubsection{Optical effects of complex magnetoelectric parameters in dielectrics}
	
	Having the refractive indices and the polarization of the propagating modes, we can examine the physical behavior brought about by the bi-isotropic constitutive relations (\ref{CRBI1A}) and (\ref{CRBI1B}) on a conventional dielectric substrate. The refractive indices (\ref{iso10A}) are written as
	\begin{equation}
		n_{\pm }=\sqrt{\mu \epsilon -\frac{\mu ^{2}(\alpha +\beta )^{2}}{4}}\mp
		\mathrm{i}\frac{\mu (\alpha +\beta )}{2}.  \label{iso19}
	\end{equation}

Let us discuss the refractive indices (\ref{iso19}) considering two
	cases: (a) $\alpha$, $\beta \in \mathbb{C}$, (b) $\alpha$, $\beta \in
	\mathbb{R}$.
	For $\alpha$ and $\beta$ complex, one can write
	\begin{equation}
		\alpha=\alpha' +\mathrm{i} \alpha'', \quad \beta=\beta' +\mathrm{i} \beta'',
		\label{alphabeta1A}
	\end{equation}
	where $\alpha'=\mathrm{Re}[\alpha]$, $\alpha''=\mathrm{Im}[\alpha]$, $\beta'=\mathrm{Re}[\beta]$ and $\beta''=\mathrm{Im}[\beta]$. The condition (\ref{SCR2}) implies
	\begin{equation}
		\alpha'=-\beta', \quad \alpha''=\beta'',
		\label{alphabeta1B}
	\end{equation}
	so that  $\alpha+\beta= 2{\mathrm{i}\alpha''}$. Therefore, \eqref{iso19} is rewritten as
	\begin{equation}
		n_{\pm }=\sqrt{\mu \epsilon +\mu ^{2}{\alpha ^{\prime \prime }}^{2}} \pm \mu
		\alpha ^{\prime \prime },  \label{isotropic-case-1-2}
	\end{equation}
	which are real, positive, and cause birefringence. Since the
	polarization modes are circularly polarized vectors, see \eqref{iso-prop-2},
	the birefringence effect can be evaluated in terms of the rotatory power, defined as
	\begin{equation}
		\delta =-\frac{[\mathrm{Re}(n_{+})-\mathrm{Re}(n_{-})]\omega }{2}\
		\label{eq:rotatory-power1A}
	\end{equation}%
	Hence, using the indices (\ref{isotropic-case-1-2}), the rotatory power is
	\begin{equation}
		\delta =- \mu \omega \alpha ^{\prime \prime }.  \label{isotropic-case-1-3}
	\end{equation}
	
	Such a birefringence effect (\ref{isotropic-case-1-3}) is a consequence of $%
	(\alpha+\beta)=2\mathrm{i}{\alpha}^{\prime\prime}$. Therefore, it only occurs when the constitutive
	parameters possess an imaginary piece. On the other hand, for $\alpha$, $\beta \in \mathbb{R}$, one has simply $\beta=-\alpha$, $\alpha''=0$, and no birefringence takes place.
	This is the case of the topological insulators bi-isotropic scenario \cite{Chang,
		Urrutia, Lakhtakia, Winder, Li,Li1}, whose constitutive
	relations are
	\begin{subequations}
		\label{iso20A}
		\begin{eqnarray}
			\mathbf{D} &=&\epsilon \mathbf{E}-\alpha _{0}\mathbf{B},  \label{iso20} \\
			\mathbf{H} &=&\frac{\mathbf{B}}{\mu }+\alpha _{0}\mathbf{E},  \label{iso21}
		\end{eqnarray}
	\end{subequations}
	with $\alpha _{0}=e^{2}/4\pi \hbar$ and $e$ being the elementary electric
	charge. For \eqref{iso20A}, one has  $(\alpha+\beta)=0$, so that no birefringence is provided. Concerning topological insulators, quantum effects of bulk interband excitations may generate strong Faraday rotation associated with a type of optical activity described by the Verdet constant \cite{Ohnoutek, Liang}. The quantum origin of this effect is not in contradiction with the classical absence of birefringence here reported.

magnetoelectric

\subsection{Bi-anisotropic media}

Bi-isotropic CRs constitute a simpler version of general bi-anisotropic linear relations, written as
\begin{subequations}
	\label{constitutive2z}
	\begin{eqnarray}
		D^{i} &=&\epsilon_{ij}E^{j}+\alpha _{ij}B^{j}, \label{constitutive2za} \\
		H^{i} &=&\beta _{ij}E^{j}+\mu^{-1}_{ij}B^{j}, \label{constitutive2zb}
	\end{eqnarray}
\end{subequations}
where $\epsilon_{ij}$ and $\mu_{ij}$ are the electrical permittivity and magnetic permeability tensors, while $\alpha_{ij}$ and $\beta_{ij}$ are magnetoelectric tensors given by $3\times 3$ complex matrices.
Bi-anisotropic ``chiral'' relations (\ref{constitutive2z}) were applied to examine energy balance issues \cite{Kamenetskii}, relativistic electron gas \cite{Carvalho}, time-dependent magneto-electric parameters \cite{Lin}, Weyl semimetals \cite{Halterman, Zu}, magnetized materials \cite{Krupka1, Krupka2}, anisotropic dispersion relations \cite{Hillion, Yakov, Damaskos}, rotation of light polarization in multiferroic materials \cite{Kurumaji}, among other systems \cite{Takahashi}.

\begin{figure}[H]
	\begin{centering}
		\includegraphics[scale=.5]{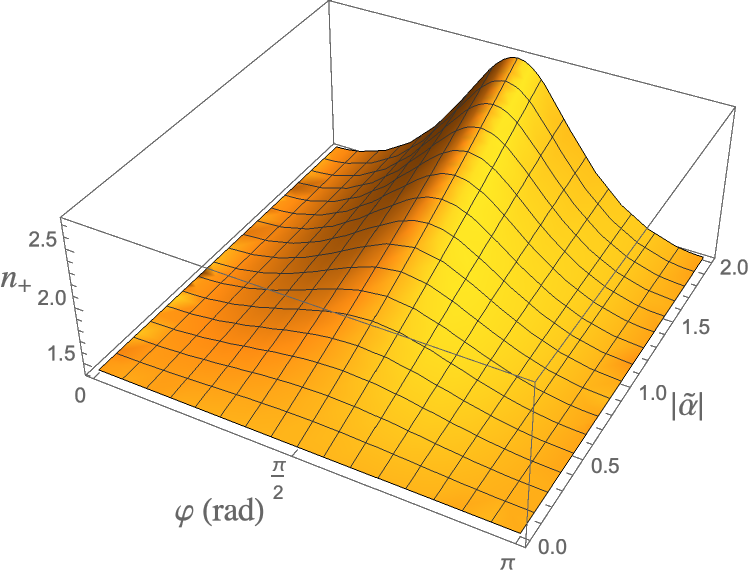}
		\par\end{centering}
	\caption{\label{plot3D-indice-de-refracao-mais-caso-simetrico} Refractive index $n_{+}$ of  \eqref{refractive-indices-symmetric-1b} with $\mu=1$, $\epsilon=2$, and $d=1$. The parameters $\epsilon$, $\mu$, and $|\tilde{\alpha}|d^{2}$ are dimensionless.}
\end{figure}

\begin{figure}[H]
	\begin{centering}
		\includegraphics[scale=.52]{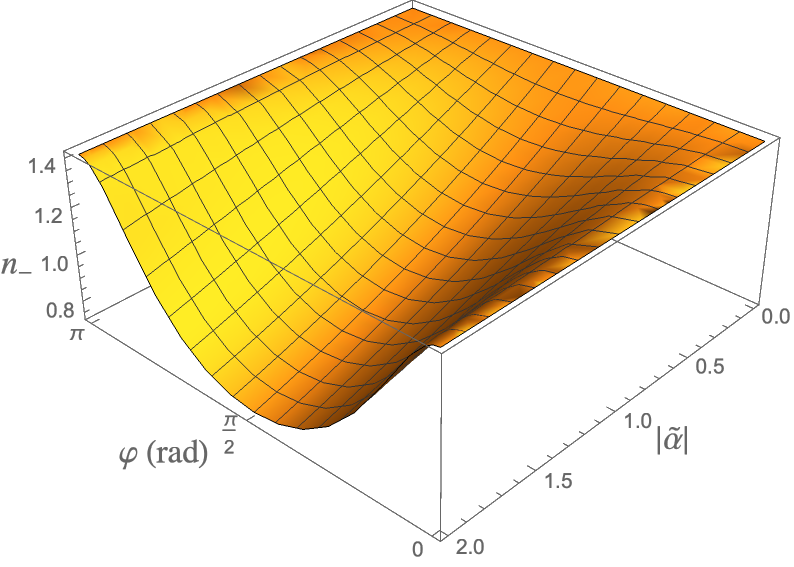}
		\par\end{centering}
	\caption{\label{plot3D-indice-de-refracao-menos-caso-simetrico-tipo-2} Refractive index $n_{-}$ of  \eqref{refractive-indices-symmetric-1b} with $\mu=1$, $\epsilon=2$, and $d=1$. The parameters $\epsilon$, $\mu$, and $|\tilde{\alpha}|d^{2}$ are dimensionless.}
\end{figure}

A recent work \cite{Pedro3} investigated how the magnetoelectric parameters of CRs (\ref{constitutive2z}) affect the propagation of electromagnetic waves and generate optical effects. The permittivity and permeability are given as diagonal and isotropic tensors, $\epsilon= \epsilon \delta_{ij}$, $\mu= \mu \delta_{ij}$, and the magnetoelectric parameters as symmetric non-diagonals tensors parameterized in terms of a fixed 3-vector, $\mathbf{d}$, that is,
\begin{equation}
	\alpha _{ij}=\tilde{\alpha}d_{i}d_{j},\quad \beta _{ij}=\tilde{\beta}%
	d_{i}d_{j}, \label{symmetric1}
\end{equation}%
so that the constitutive relations (\ref{constitutive2}) take the form,
\begin{subequations}
	\label{constitutive-relations-symmetric-1B}
	\begin{align}
		\mathbf{D}& =\epsilon \mathbf{E}+\tilde{\alpha}\mathbf{d}(\mathbf{d}\cdot
		\mathbf{B}), \label{symmetric2-1} \\
		\mathbf{H}& =\frac{1}{\mu }\mathbf{B}+\tilde{\beta}\mathbf{d}(\mathbf{d}%
		\cdot \mathbf{E}). \label{symmetric2A}
	\end{align}%
\end{subequations}
In such a context, the Maxwell equations yield the wave equation (\ref{WE1A}) with the effective permittivity tensor,
\begin{equation} \label{symmetric3}
	\bar{\epsilon}_{ij}=\epsilon\delta_{ij}-\frac{1}{\omega}\left( \tilde{\beta}\varepsilon_{imn}k_{m}d_{n}d_ {j}+ \tilde{\alpha}\varepsilon_{amj} d_{i}d_{a}k_{m}
	\right).
\end{equation}
Replacing relations (\ref{symmetric1}) in the permittivity tensor (\ref{symmetric3}), ones writes
\begin{align}  \label{symmetric3}
	\bar{\epsilon}_{ij}=\epsilon\delta_{ij}-\frac{1}{\omega}\left( \tilde{\beta%
	}   \varepsilon_{imn}      k_{m}d_{n}d_{j}+ \tilde{\alpha}    \varepsilon_{amj}   d_{i}d_{a}k_{m}%
	\right),
\end{align}
in such a way the tensor $M_{ij}$, \eqref{ex12}, provides
\begin{align}
	M = \mathcal{N} - \mu (\mathcal{D} + \mathcal{E}) , \label{symmetric4}
\end{align}
with $\mathcal{N}$ given by \eqref{Nij1}, while the magnetoelectric  contributions are displayed as
\begin{subequations}
	\label{compact-matrices-symmetric-case-1}
	\begin{align}
		\mathcal{D} &= -(\tilde{\alpha}-\tilde{\beta})  \mathrm{diag} \left(D_{1}, D_{2}, D_{3} \right),  \\
		\mathcal{E} &= \begin{pmatrix}
			0 & \epsilon_{12} & \epsilon_{13} \\
			\epsilon_{21} & 0 & \epsilon_{23} \\
			\epsilon_{31} & \epsilon_{32} & 0
		\end{pmatrix} ,
	\end{align}
	where
	\begin{align}
		D_{1}&= d_{1} (d_{2} n_{3} -d_{3} n_{2}), \\
		D_{2}&=  d_{2} (d_{3} n_{1} - d_{1} n_{3}), \\
		D_{3} &=  d_{3} (d_{1} n_{2} - d_{2} n_{1} ) ,
	\end{align}
\end{subequations}
and
\begin{subequations}
	\label{matrix-symmetric-case-extra-0}
	\begin{align}
		\epsilon_{12} &= -\tilde{\beta}d_{2} (d_{3}n_{2}-d_{2}n_{3}) + \tilde{\alpha}%
		d_{1}(d_{1}n_{3}-d_{3}n_{1}) ,  \label{symmetric-case-extra-1} \\
		\epsilon_{13} &= -\tilde{\beta}d_{3} (d_{3}n_{2} - d_{2}n_{3}) + \tilde{%
			\alpha} d_{1} (d_{2}n_{1}-d_{1}n_{2}) ,  \label{symmetric-case-extra-2} \\
		\epsilon_{21} &= -\tilde{\beta}d_{1} (d_{1}n_{3} -d_{3}n_{1}) + \tilde{\alpha%
		}d_{2} (d_{3}n_{2} - d_{2}n_{3}),  \label{symmetric-case-extra-3} \\
		\epsilon_{23} &= - \tilde{\beta} d_{3} (d_{1}n_{3} - d_{3}n_{1}) + \tilde{%
			\alpha}d_{2} (d_{2}n_{1} - d_{1}n_{2}) ,  \label{symmetric-case-extra-4} \\
		\epsilon_{31} &= - \tilde{\beta}d_{1} (d_{2} n_{1} - d_{1} n_{2} ) + \tilde{%
			\alpha} d_{3} (d_{3} n_{2} -d_{2}n_{3}),  \label{symmetric-case-extra-5} \\
		\epsilon_{32} &= -\tilde{\beta} d_{2} (d_{2} n_{1} - d_{1}n_{2}) + \tilde{%
			\alpha} d_{3} (d_{1}n_{3} - d_{3} n_{1} ) .  \label{symmetric-case-extra-6}
	\end{align}
Evaluating $\mathrm{det}[M_{ij}]=0$, we obtain the dispersion relation,
\end{subequations}
\begin{align}  \label{symmetric5}
	{\epsilon} \left(n^{2}-\mu\epsilon \right)^{2} +
	\tilde{\alpha} \tilde{\beta} \mu \left[ \mu {\epsilon} d^{2} - (%
	\mathbf{n}\cdot \mathbf{d} )^{2} \right]
	{\left( \mathbf{d}\times \mathbf{n}\right) ^{2}} =0 ,
\end{align}
{ where $\left( \mathbf{d}\times \mathbf{n}\right) ^{2}\equiv d^2\mathbf{n}^2 -\left( \mathbf{d}\cdot\mathbf{n}\right) ^{2}$.}
The relation (\ref{SCR2}) provides $\tilde{\alpha}
\tilde{\beta}=-|\tilde{\alpha}|^{2}$.
Furthermore, implementing $\mathbf{n}\cdot \mathbf{d}=n
d\cos \varphi$, \eqref{symmetric5} provides the following refractive indices:
\begin{equation}
	n_{\pm }^{2}=\frac{1}{s}\left[ N\pm \mu |\tilde{\alpha}|d^{2}\sin
	^{2}\varphi \sqrt{\mu \epsilon +\frac{\mu ^{2}|\tilde{\alpha}|^{2}d^{4}}{4}}%
	\right] , \label{refractive-indices-symmetric-1}
\end{equation}%
where we use $\mathbf{n}\cdot \mathbf{d}=nd\cos \varphi$. Such indices are real, positive, and can also be read as,
\begin{equation}
	n_{\pm }=\sqrt{\frac{N+\mu \epsilon \sqrt{s}}{2s}}\pm \sqrt{%
		\frac{N-\mu \epsilon \sqrt{s}}{2s}}, \label{refractive-indices-symmetric-1b}
\end{equation}%
with
\begin{subequations}
	\label{refractive-indices-symmetric-2}
	\begin{align}
		N& =\mu \epsilon +\frac{\mu ^{2}|\tilde{\alpha}|^{2}d^{4}}{2}\sin
		^{2}\varphi . \label{refractive-indices-symmetric-4} \\
		s& =1+\frac{\mu }{\epsilon }|\tilde{\alpha}|^{2}d^{4}\sin ^{2}\varphi \cos
		^{2}\varphi . \label{refractive-indices-symmetric-3}
	\end{align}
\end{subequations}
The indices (\ref{refractive-indices-symmetric-1b}) hold for any propagation direction in relation to the vector $\mathbf{d}$, validity which is assured in terms of the angle $\varphi$ (the relative angle between the vector ${\bf{d}}$ and the propagation direction). The behavior of $n_{\pm}$ as a function of $\varphi \in [0, \pi]$ and the dimensionless parameter $|\tilde{\alpha}| \in [0,1]$ is shown in Figs.~\ref{plot3D-indice-de-refracao-mais-caso-simetrico} and \ref{plot3D-indice-de-refracao-menos-caso-simetrico-tipo-2}.

In the following, we address the propagating modes and examine birefringence effects {for some specific propagation directions.}

\subsection{\label{section-propagation-modes-symmetric-case}Propagation modes}

We firstly examine the propagating modes for the case the vectors $\mathbf{n}$ and $\mathbf{d}$ are parallel, setting 	$\mathbf{d}=(0,0,d)$ and $\mathbf{n}=(0,0,n)$, for which \eqref{refractive-indices-symmetric-1} yields $n_{\pm}=\sqrt{\mu\epsilon}$, the refractive index of the usual scenario. In this scenario, equation $M_{ij}E^{j}=0$ provides generic orthogonal modes
\begin{equation}
	\mathbf{E}=%
	\begin{pmatrix}
		E_{x} \\
		E_{y} \\
		0%
	\end{pmatrix}%
	,  \label{propagating-symmetric-6}
\end{equation}%
representing a transversal wave with undefined polarization (linear, circular, or elliptical).
A different scenario is the $\mathbf{d}$-transversal configuration, defined by 
\begin{equation}
	\mathbf{d}=(d_{1},d_{2},0),  \label{d-T}
\end{equation}%
for $\mathbf{n}=(0,0, n)$, which implies $s=1$ and $N=\mu {\epsilon }+|\tilde{\alpha}|^{2}\mu ^{2}d^{4}/2$. Thus,
one obtains
\begin{equation}
	n_{\pm }^{2}=\mu \epsilon +\frac{\mu ^{2}|\tilde{\alpha}|^{2}d^{4}}{2}\pm
	\mu |\tilde{\alpha}|d^{2}\sqrt{\mu \epsilon +\frac{\mu ^{2}|\tilde{\alpha}%
			|^{2}d^{4}}{4}},  \label{symmetric6T}
\end{equation}%
and
\begin{equation}
	n_{\pm }=\sqrt{\mu \epsilon +\frac{\mu ^{2}|\tilde{\alpha}|^{2}d^{4}}{4}}\pm
	\frac{\mu |\tilde{\alpha}|d^{2}}{2}.  \label{symmetric6T2}
\end{equation}
By using Eq. (\ref{symmetric6T2}) we rewrite Eq. (\ref{symmetric6T}) as
\begin{equation}
	n_{\pm }^{2}=\mu \epsilon \pm \mu |\tilde{\alpha}| d^{2} n_\pm.
	\label{symmetric6T0}
\end{equation}%
The matrix (\ref{symmetric4}) now reads
\begin{equation}
	M=
	\begin{pmatrix}
		(\pm \mu |\tilde{\alpha}| d^{2}+\Omega )n_{\pm } & -\mu n_{\pm
		}(\tilde{\beta}d_{2}^{2}+\tilde{\alpha}d_{1}^{2}) & 0 \\
		\mu n_{\pm }(\tilde{\beta}d_{1}^{2}+\tilde{\alpha}d_{2}^{2}) & (
		\pm \mu |\tilde{\alpha}| d^{2}-\Omega )n_{\pm } & 0 \\
		0 & 0 & -\mu \epsilon%
	\end{pmatrix}%
	,
\end{equation}
with
\begin{align}
	\Omega & =\mu (\tilde{\alpha}-\tilde{\beta})d_{1}d_{2}=2\mu \tilde{\alpha}%
	^{\prime }d_{1}d_{2},
\end{align}%
written from \eqref{symmetric6T0}. The condition $M_{ij}E^{j}=0$ yields
\begin{equation}
	\mathbf{E}_{\pm }=E_{0}%
	\begin{pmatrix}
		1 \\
		\displaystyle\frac{\mu \left( \tilde{\beta}d_{1}^{2}+\tilde{\alpha}%
			d_{2}^{2}\right) }{\Omega \mp \mu |\tilde{\alpha}|d^{2}} \\
		0%
	\end{pmatrix},
\end{equation}%
with an appropriately chosen amplitude $E_{0}$.
For
$d_{1}=0$, we achieve
\begin{equation}
	\mathbf{E}_{\pm }=E_{0}%
	\begin{pmatrix}
		1 \\
		\displaystyle\mp \frac{\tilde{\alpha}}{|\tilde{\alpha}|} \\
		0%
	\end{pmatrix}%
	=\frac{1}{\sqrt{2}}%
	\begin{pmatrix}
		1 \\
		\displaystyle\mp \frac{\tilde{\alpha}^{\prime }+\mathrm{i}\tilde{\alpha}%
			^{\prime \prime }}{|\tilde{\alpha}|} \\
		0%
	\end{pmatrix}%
	,  \label{propagating-symmetric-6B-1}
\end{equation}%
which may represent,
\begin{itemize} 
\item linear polarizations for $\tilde{\alpha}^{\prime \prime
}=0$,
\item circular polarizations for $\tilde{\alpha}^{\prime }=0$.
\end{itemize}
Given the two real refractive indices given in \eqref{symmetric6T2},
a birefringent scenario is established. As the vectors
(\ref{propagating-symmetric-6B-1}) do not represent RCP or LCP
modes, one does not have circular birefringence, that is, the propagation can not be suitably described
in terms of the rotatory power \eqref{eq:rotatory-power1A}.
Rather, it can be characterized in terms of the phase
shift arising from the distinct phase velocities of the propagating modes,
given by
\begin{equation}
	\Delta =\frac{2\pi }{\lambda _{0}}l(n_{+}-n_{-}),  \label{phase-shift-0}
\end{equation}%
where $\lambda _{0}$ is the vacuum wavelength of incident light, $l$ is the thickness of the medium or the distance
traveled by the wave, $n_{+}$ and $n_{-}$	are the refractive indices of the medium. Note that this is the same expression that controls the phase shift caused
by ``retarders'' (for details, see Chapter 8 of Ref. \cite{Hecht}). Replacing the indices (\ref{symmetric6T2}) in \eqref{phase-shift-0}, one obtains 
\begin{equation}
	\frac{\Delta }{l}=\frac{2\pi }{\lambda _{0}}\mu |\tilde{\alpha}|d^{2},
	\label{phase-shift-1}
\end{equation}
which represents the phase shift per unit length. As the phase shift depends on the modulus of $\tilde{\alpha}$, the birefringence now takes place for both real and imaginary magnetoelectric parameters. This is a difference in relation to the bi-isotropic case of \eqref{constitutive-relations-biisotropic-1}, in which the birefringence only occurs for imaginary parameters, as shown in \eqref{isotropic-case-1-3}. For further details, see Ref. \cite{Pedro3}.

\section{\label{section3}Propagation behavior under magnetic conductivity}

In this section, we investigate how the magnetic conductivity affects electromagnetic wave propagation in a dielectric substrate, employing the usual classical framework. In this sense, we consider the linear magnetic current,
\begin{equation}
J^{i}={\sigma }_{ij}^{B}B^{j},  \label{eq20}
\end{equation}%
where ${\sigma}_{ij}^{B}$ is the conductivity tensor, the effective manifestation of the medium's response to the magnetic field in the context of the CME.

Research on the CME
\cite{Kharzeev1,Fukushima,Kharzeev2,Qiu} have reported the
generation of a current density entailed by a magnetic field,
\begin{equation}
	J_{\mathrm{CME}}^{i}={\frac{e^{2}}{{4{\pi }^{2}}}}({\Delta }\mu )B^{i}\equiv \Sigma B^i\,,  \label{eq70}
\end{equation}%
where $e$ is the fermion charge, $\mathbf{B}$ the applied magnetic field,
${\Delta }\mu \equiv {\mu }_{R}-\mu _{L}$ is also known as the chiral
chemical potential, and 
\begin{equation}
	{\Sigma}={\frac{e^{2}}{{4{\pi }^{2}}}}{\Delta }\mu\,.  \label{eq74}
\end{equation}%
is the chiral magnetic conductivity~\cite{Maxim,Leite,Dvornikov,Bubnov,Akamatsu,Maxim1,Burkov}. The general conductivity of \eqref{eq70} can recover the isotropic diagonal magnetic conductivity of the CME, that is,	
	\begin{equation}
		{\sigma}_{ij}^{B}={\Sigma \delta }_{ij}\,. \label{eq72}
	\end{equation}
The general magnetic conductivity ${\sigma }_{ij}^{B}$ can be split as 
	\begin{equation}
		{\sigma}_{ij}^{B}={\Sigma \delta}_{ij}+{\Sigma}_{ij}\,,  \label{eq69}
	\end{equation}%
where the isotropic part fulfills ${\Sigma}=(1/3) Tr({\sigma}_{ij}^{B})$ and ${\Sigma }_{ij}$
	stands for all off-diagonal anisotropic components of the matrix ${\sigma }_{ij}^{B}$.  Investigations discussing the off-diagonal anisotropic magnetic current,
	\begin{equation}
		J^{i}={\Sigma}_{ij}B^{j},  \label{eq75}
	\end{equation}
have also been reported \cite{Pedro1,PedroPRB2024A,PedroPRB2024B,PedroPRB}, where its consequences on electromagnetic propagation and optical effects in dielectrics were discussed. Further, antisymmetric magnetic conductivity has found realization in systems of condensed matter \cite{Kaushik2}. Although the usual CME is associated with isotropic conductivity (\ref{eq72}), the off-diagonal components are also associated with CME in Weyl semimetals.

\subsection{Dispersion relation and refractive indices}

Using the generalized Ohm's law of \eqref{eq20} and conventional isotropic
constitutive relations, $D^{i}=\epsilon {\delta }_{ij}E^{j}$,
$H^{i}=\mu^{-1}{\delta}_{ij}B^{j}$, in the Maxwell equations,
\eqref{eq10} keeps its general form:
\begin{subequations}
\begin{equation}
\left[ \mathbf{k}\times \mathbf{k}\times {\mathbf{E}}\right] ^{i}+{\omega }%
^{2}\mu \bar{\epsilon}_{ij}(\omega )E^{j}=0\,,  \label{eq21}
\end{equation}%
where
\begin{equation}
{\bar{\epsilon}}_{ij}(\omega )=\left( \epsilon +\mathrm{i}{\frac{\sigma }{\omega }}%
\right) {\delta }_{ij}+{\frac{\mathrm{i}}{{\omega }^{2}}}({\sigma }^{B})_{ia}{%
\varepsilon }_{ajb}k_{b}\,,  \label{eq22}
\end{equation}%
\end{subequations}
defines the frequency-dependent extended permittivity tensor.
Equation~(\ref{eq21}) implies
\begin{equation}
	\left[n^2{\delta }_{ij}-n^{i}n^{j}-\mu \bar{{\epsilon }}_{ij}%
	\right] E^{j}=0\,.  \label{eq30}
\end{equation}

The latter can also be cast into the form
\begin{subequations}
\begin{equation}
M_{ij}E^{j}=0\,,  \label{MijE}
\end{equation}%
where the tensor $M_{ij}$ reads
\begin{equation}
M_{ij}=n^{2}{\delta }_{ij}-n_{i}n_{j}-\mu \bar{{\epsilon }}_{ij}\,, \label{eq36B}
\end{equation}%
\end{subequations}
and $\bar{{\epsilon }}_{ij}$ is given by \eqref{eq22}. Requiring $\mathrm{det}[M_{ij}]=0$ provides the dispersion relations and refractive indices, which allows us to determine the corresponding polarizations for the propagating modes.

In the following, we will discard refractive indices with negative real parts
associated with frequencies that have the same property. Sophisticated composites of
different materials can be designed that have negative permittivity and permeability.
These are called metamaterials \cite{Kshetrimayum} and the real parts of their refractive indices must be
endowed with a $\pm$ sign according to the materials at the interface:
material-material (+), material-metamaterial ($-$), metamaterial-metamaterial (+).
On the contrary, negative refractive indices are not known to occur in crystals found
in nature, which our focus is on in this paper.

We consider now the case of an isotropic magnetic conductivity,
represented by the diagonal isotropic matrix of \eqref{eq72}. By implementing the latter in \eqref{eq22}, one obtains
\begin{equation}
{\bar{\epsilon}}_{ij}(\omega )=\left( \epsilon +\mathrm{i}{\frac{\sigma }{\omega }}%
\right) {\delta }_{ij} + {\frac{{\mathrm{i}\Sigma }}{{\omega }^{2}}}{\epsilon }%
_{ijb}k_{b}\,.  \label{eq61a}
\end{equation}

\begin{figure}[h]
\begin{centering}
\includegraphics[scale=0.69]{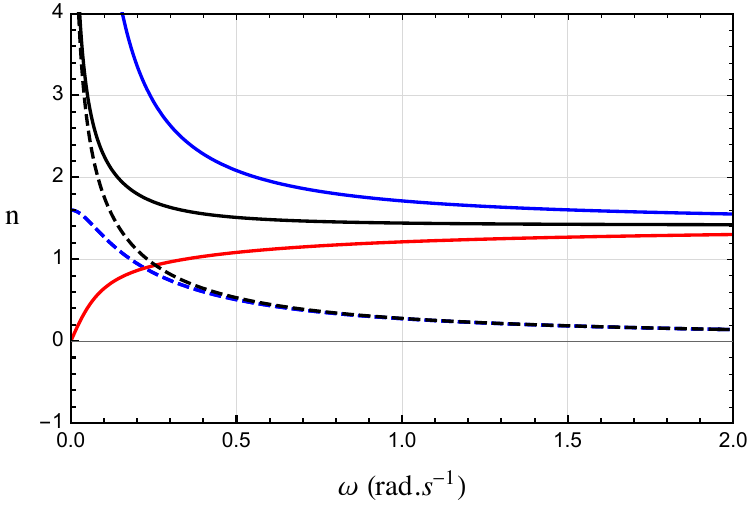}
\par\end{centering}
\caption{\label{plot-indices-mais-menos-isotropic-case} Refractive indices $n_{\pm}$ of \eqref{eq63b} in terms of $\omega$. Solid (dashed) lines indicate the real (imaginary) parts of $n_{+}$ (blue), $n_{-}$ (red). The usual case $n \, (\Sigma=0)$ is represented by the black lines. The dashed blue and red curves lie on top of each other, indicating the same absorption behavior for both indices. Here we have used $\mu=1$, $\epsilon=2$, $\sigma=0.8$~$\mathrm{s}^{-1}$, $\Sigma=0.5$ $\mathrm{s}^{-1}$.}
\end{figure}

The second term on the right-hand side of \eqref{eq61a} is an antisymmetric permittivity part that stands for the contribution of the magnetic conductivity. Such an antisymmetric piece is explicitly written in the matrix form in Eq. (\ref{eq62}),
\begin{equation}
\label{eq62}
[M_{ij}]=\tilde{\mathcal{N}}-i\frac{\mu\Sigma}{\omega} \begin{pmatrix}
		0 & n_{3} & - n_{2} \\
		- n_{3} & 0 & n_{1} \\
		n_{2} & - n_{1} & 0
	\end{pmatrix}  ,
	\end{equation} 
where
\begin{equation}
\label{eq66-2}
\tilde{\mathcal{N}}=\begin{pmatrix}
		n_2^2+n_3^2-\mu \tilde{\epsilon} & -n_1n_2 & -n_1n_3 \\
		-n_1n_2 & n_1^2+n_3^2-\mu \tilde{\epsilon} & -n_2n_3 \\
		-n_1n_3 & -n_2n_3 & n_1^2+n_2^2-\mu \tilde{\epsilon} \\
	\end{pmatrix}\, .
\end{equation}
Here, we use the compact notation, $\tilde{\epsilon}=\epsilon+ i (\sigma/\omega)$, to include the Ohmic conductivity. To obtain the dispersion relation (in terms of the refractive index), we impose $\det[M_{ij}]=0$, yielding
\begin{subequations}
\begin{align}
\label{eq63}
n_{\pm}^{2}&=\mu\epsilon+ i \frac{\mu \sigma}{\omega}+\frac{\mu^{2}\Sigma^{2}}{2\omega^{2}} \pm \frac{\mu\Sigma}{\omega} \,
\sqrt{\mu\epsilon + i  \frac{ \mu \sigma}{\omega}+\left(\frac{\mu\Sigma}{2\omega}\right)^2}, 
\end{align}
which provides,
\begin{align}
n_{\pm} &= \sqrt{ \mu \epsilon + i \frac{\mu\sigma}{\omega} + \left(\frac{ \mu \Sigma}{2\omega} \right)^{2}} \pm \frac{\mu \Sigma}{2\omega}. \label{eq63-1-1}
\end{align}
\end{subequations}
These refractive indices are also written as
\begin{subequations}
\begin{align}
n_{\pm} &= \left(I_{+} + i I_{-} \right) \pm \frac{\mu\Sigma}{2\omega} , \label{eq63b}
\end{align}
with
\begin{align}
I_{\pm} &= \frac{\hat{I}_{\pm}}{\sqrt{2}} \sqrt{\mu\epsilon + \left(\frac{\mu\Sigma}{2\omega}\right)^{2}} \; ,
\label{isotropic-case-3}
\\
\hat{I}_{\pm} &=\sqrt{  \sqrt{ 1 + \left( \frac{\mu (\sigma/\omega)}{\mu\epsilon+ \mu^{2}\Sigma^{2} /(4\omega^{2})} \right)^{2}} \pm 1} \; , \label{isotropic-case-3-1}
\end{align}
\end{subequations}
where the imaginary piece is now explicitly separated in terms of $I_{-}$. Equation~(\ref{eq63b}) contains two distinct refractive indices for each propagation direction, which is a signature of anisotropy and yields manifestation of birefringence. As the dielectric substrate is isotropic, that is, it is described by an isotropic permittivity tensor, $\epsilon_{ij}= \epsilon {\delta}_{ij}$, the anisotropy effect stems from the way the magnetic conductivity is coupled to the electromagnetic fields. The expressions (\ref{eq63b}) reveal how the chiral	conductivity affects the refractive index of a dielectric medium with Ohmic conductivity ($\sigma \neq 0)$, providing two very distinct behaviors at the low-frequency regime (near the origin). The latter is depicted in Fig. \ref{plot-indices-mais-menos-isotropic-case}, where $n_{\pm}$ are given in terms of frequency $\omega$. While the real piece of $n_+$ presents an analog behavior to the usual index $n=\sqrt{\mu \epsilon}$ ($\Sigma=0$), the real piece of the index $n_-$ goes to zero at the origin, a bold difference in relation the divergent behavior of $\mathrm{Re}(n_+)$ at $\omega=0$. On the other hand, the imaginary pieces of both indices are equal, $\mathrm{Im}(n_+)=\mathrm{Im}(n_-)=I_-$, equality represented in terms of the single dashed blue line in Fig. \ref{plot-indices-mais-menos-isotropic-case}.

Considering the case of a dielectric medium with zero Ohmic conductivity, $\sigma =0$, the relation (\ref{eq63-1-1}) yields two distinct real and frequency-dependent refractive indices,
\begin{align}
n_{\pm} &= \sqrt{ \mu \epsilon + \left(\frac{ \mu \Sigma}{2\omega} \right)^{2} } \pm \frac{\mu \Sigma}{2\omega} ,
\label{n23-1}
\end{align}
describing a dispersive and non-absorbing matter, which allows free attenuation electromagnetic propagation. Such a medium has the properties of a biaxial crystal, thus endowed with anisotropy and birefringence.

It is relevant to emphasize that the magnetic conductivity, $\Sigma$, only implies an absorbing behavior for the
medium when it occurs on a dielectric substrate endowed with Ohmic conductivity, that is, $\sigma \neq 0,\Sigma\neq 0$, as shown in the complex refractive index of Eqs. (\ref{eq63-1-1}) and (\ref{eq63b}). On the other hand, when the magnetic current is realized on a lossless dielectric substrate $(\sigma =0, \Sigma \neq 0)$, the medium behaves as a dispersive non-absorbing dielectric, see Eq.(\ref{n23-1}), despite the presence of chiral conductivity. This happens because $I_{-}=0$ when $\sigma=0$.

\subsection{Propagation modes and birefringence}

The propagating modes describe the polarization of the electromagnetic waves that travel through the medium with phase velocities associated with the refractive indices, $c/n_{\pm}$. For each refractive index, $n_{\pm}$, there will be an electromagnetic wave whose polarization (electric field) vector can be determined from the equation $M_{ij}E^{j}=0$. Replacing \eqref{eq63-1-1} in the matrix $M_{ij}$ of \eqref{eq62}, the equation $M_{ij}E^{j}=0$ provides the following propagation modes:
\begin{equation}
\label{eigenvector2}
\mathbf{E}_{\pm}=\frac{1}{n\sqrt{2(n_1^2+n_3^2)}}
\begin{pmatrix}
n n_3 \pm \mathrm{i}n_1 n_2 \\
\mp\mathrm{i}(n_1^2+n_3^2) \\
\pm\mathrm{i}n_2 n_3-n n_1 \\
\end{pmatrix} \; .
\end{equation}

To get more physical insights on the polarization of the propagating modes, let us consider, without loss of generality, the propagation direction along the $z$-axis, i.e., ${\bf{n}}= (0, 0, n)$. Thus, the $M_{ij}$ matrix becomes

\begin{align}
[M_{ij} ]&=
\begin{pmatrix}
n^{2} - \mu \tilde{\epsilon} & - i \mu \frac{\Sigma}{\omega} n & 0
\\
\\
+i\mu  \frac{\Sigma}{\omega} n & n^{2} - \mu \tilde{\epsilon} & 0
\\
\\
0 & 0 & - \mu \tilde{\epsilon}
\end{pmatrix} .
\label{propagating-modes-isotropic-1}
\end{align}
From \eqref{eq63-1-1}, one writes
\begin{align}
n^{2}_{\pm} - \mu \tilde{\epsilon}= \pm \, n_{\pm} \, \frac{\mu \Sigma}{\omega} \; .
\label{propagating-modes-isotropic-2}
\end{align}
Now by implementing \eqref{propagating-modes-isotropic-2} in \eqref{propagating-modes-isotropic-1} and solving $M_{ij}E^{j}=0$, one finds
\begin{align}
{\bf{E}}_{\pm} &= \frac{1}{\sqrt{2}} \begin{pmatrix}
1\\
\mp \, i\\
0
\end{pmatrix} \; , \label{propagating-modes-isotropic-3}
\end{align}
which represent right-handed (RCP) and left-handed (LCP) circularly polarized vectors, respectively. Details about these modes are better explained in Refs.~\cite{Pedro1,PedroPRB2024A}.

A consequence of the optical activity of the medium (birefringence) is the polarization rotation of an incident linearly polarized wave. Such an effect occurs because the two circularly polarized propagating modes of opposite chiralities, see \eqref{propagating-modes-isotropic-3}, travel through the birefringent medium with different phase velocities, $c/ \mathrm{Re}[n_{\pm}]$. Hence, the rotation of the initial linear polarization of the incident light is measured using the specific rotatory power $\delta$ \cite{Pedro2}, which measures the rotation of the polarization of linearly polarized light per unit traversed length in the medium, given by
\begin{align}
\delta = -\frac{\omega}{2} \left( \mathrm{Re}[n_{+}]- \mathrm{Re}[n_{-}] \right) . \label{rotatory-power-1}
\end{align}
Using \eqref{eq63b} in \eqref{rotatory-power-1}, we find
\begin{equation}
\label{eq:rotatory-power}
\delta=-\frac{\mu\Sigma}{2}\,,
\end{equation}
thus providing a frequency-independent specific rotatory power due to the chiral magnetic
conductivity $\Sigma$, which measures the rotation (per unit length) of the polarization plane of a linearly polarized wave.

The general behavior found for this configuration is very interesting. The magnetic
conductivity $\sigma^B_{ij}$ is odd under parity inversion \cite{Pedro1}. Although parity violation does
not appear in a single refractive index of \eqref{n23-1}, which behaves isotropically (it does not depend on the propagation direction), it arises in the distinct propagation properties of left- and right-handed polarized electromagnetic waves, yielding the birefringence. Such a property takes place in an initially isotropic dielectric substrate which becomes anisotropic in the presence of the magnetic conductivity, exhibiting a biaxial crystal structure \cite{PedroPRB2024A}. Indeed, for the wave propagation  in general direction, ${\bf{n}}={(k_1,k_2,k_3)}/\omega$, and $\sigma=0$, the permittivity tensor (\ref{eq61a}) reads
	\begin{align}
		\left[ \bar{\epsilon}_{ij} \right] &= \begin{pmatrix}
			\epsilon && \frac{i\Sigma}{\omega^{2}} k_{3} && - \frac{i \Sigma}{\omega^{2}} k_{2} \\
			\\
			-\frac{i \Sigma}{\omega^{2}} k_{3} && \epsilon && \frac{i\Sigma}{\omega^{2}} k_{1} \\
			\\
			\frac{i \Sigma}{\omega^{2}} k _{2} && - \frac{i \Sigma}{\omega^{2}} k_{1} && \epsilon
		\end{pmatrix},  \label{crystal-symmetry-23}
	\end{align} 
which possesses three distinct eigenvalues:
\begin{equation}
\epsilon,  \epsilon_{+}=\epsilon + \Sigma k /\omega^2, \epsilon_{-}=\epsilon - \Sigma k /\omega^2,
\end{equation}
identified as the principal permittivity values, with $k=\sqrt{{\bf{k}}^{2}}$. Hence,  after being reduced to its principal axes system, the permittivity (\ref{crystal-symmetry-23}) represents a biaxial crystal, which embraces triclinic, monoclinic, and orthorhombic systems \cite{PedroPRB2024A}. This is a feature of chiral matter with anomalous Hall effect \cite{Qiu} described by the spacelike Maxwell-Carroll-Field-Jackiw electrodynamics \cite{Pedro2}. 

Another relevant feature of the permittivity (\ref{crystal-symmetry-23}) is its hermiticity, which assures lossless propagation (free of dissipation). This and other interesting properties of dielectrics endowed with magnetic conductivity were examined in Refs. \cite{PedroPRB2024A,PedroPRB2024B}.

\section{\label{RP_reversion}Rotatory power reversion in a bi-isotropic dielectric with magnetic current}

We now discuss the propagation of electromagnetic waves in a dielectric medium endowed with two distinct sources of chirality, bi-isotropic constitutive relations, and magnetic conductivity, whose conjunction brings about the effect of rotatory power sign reversion \cite{PedroPRB}, as will be described below.  

The dielectric medium is ruled by the following constitutive relations:
\begin{subequations}
	\label{constitutive-relations-biisotropic-1}
	\begin{align}
		\mathbf{D}& =\epsilon \mathbf{E}+\alpha \mathbf{B},
		\label{constitutive-iso1D} \\
		\mathbf{H}& =\frac{1}{\mu }\mathbf{B}+\beta \mathbf{E},
		\label{constitutive-iso2D} \\
		\mathbf{J}& ={\Sigma}\mathbf{B},
		\label{constitutive-iso3D}
	\end{align}%
\end{subequations}
with $\alpha$, $\beta \in \mathbb{C}$, and $\Sigma \in \mathbb{R}$ representing the isotropic magnetic conductivity. Starting from the Maxwell equations and the relations (\ref{constitutive-relations-biisotropic-1}), one obtains the permittivity effective tensor
\begin{equation}
	\bar{\epsilon}_{ij}= \epsilon \delta _{ij}+\left(\alpha +\beta +\mathrm{i} \frac{\Sigma}{\omega} \right) \epsilon _{ijm}n_{m},
	\label{iso4}
\end{equation} 
where the last term in RHS represents the ``magnetic-electric" contribution to the medium permittivity. In
this case, the tensor $M_{ij}$ (\ref{ex12}) has the form
	\begin{align}
		[M_{ij}] = \mathcal{N} - \mu \left( \alpha+\beta+\mathrm{i}\frac{\Sigma}{\omega}\right)  \begin{pmatrix}
			0&  n_{3} &- n_{2} \\
			-n_{3} & 0 & n_{1} \\
			n_{2} &  -n_{1} & 0
		\end{pmatrix}   , \label{m-matrix-full-isotropic-1}
	\end{align}
with $\mathcal{N}$ defined in Eq. (\ref{Nij1}). The condition $\mathrm{det}[M_{ij}]=0$ yields
\begin{equation}
	n^{4}-n^{2} \left[ 2\mu {\epsilon}  -\mu ^{2} \left(\alpha +\beta+\mathrm{i}\frac{\Sigma}{\omega} \right)^{2}\right] +\mu ^{2} {\epsilon}^{2} =0  .  \label{iso6}
\end{equation}%
Solving for $n$, we obtain the following refractive indices
\begin{equation}
	n_{\pm }=\sqrt{\mu {\epsilon} -Z} \mp \mathrm{i}\sqrt{Z}  ,
	\label{iso10}
\end{equation}
with
\begin{align}
	Z&=\frac{\mu ^{2}}{4} \left(\alpha +\beta+\mathrm{i}\frac{\Sigma}{\omega} \right)^{2} .  \label{iso-7-1}
\end{align}

As we have started with isotropic tensors, $\epsilon \delta
_{ij} $, $\mu ^{-1}\delta _{ij}$, $\alpha \delta _{ij}$, $\beta \delta _{ij}$, any arising birefringence effects stem from the way magnetoelectric and magnetic conductivity are coupled to the fields in the constitutive relations (\ref{constitutive-relations-biisotropic-1}).

For achieving the propagating modes, the refractive indices (\ref{iso10}) 
\begin{align}
	n_{\pm}^{2} = \mu{\epsilon} \mp 2 \mathrm{i}\sqrt{Z} n_{\pm}, \label{full-iso-propagation-1}
\end{align} 
are replaced in the matrix (\ref{m-matrix-full-isotropic-1}) and \eqref{ex11} yields
\begin{align}
	{\bf{E}}_{\pm} &= \frac{1}{\sqrt{2}n\sqrt{n^{2}-n_{1}^{2}}} \begin{pmatrix}
		n^{2}-n_{1}^{2} \\
		\mp \mathrm{i} n n_{3} -n_{1}n_{2} \\
		\pm \mathrm{i} n n_{2} - n_{1} n_{3} 
	\end{pmatrix} . \label{full-iso-propagation-2}
\end{align}
For specific propagation along the $z$-axis, ${\bf{n}}=(0,0,n)$, the modes (\ref{full-iso-propagation-2}) reduce to right-handed ($\hat{\bf{E}}_{+}$) and left-handed circular ($\hat{\bf{E}}_{-}$) polarization vectors, 
\begin{align}
	{\bf{E}}_{\pm} &=\frac{1}{\sqrt{2}} \begin{pmatrix}
		1\\
		\mp \mathrm{i} \\
		0
	\end{pmatrix}, \label{full-iso-propagation-3}
\end{align} 
which yields circular birefringence. Considering the complex character of the parameters $\alpha$, $\beta$, and the relations (\ref{alphabeta1A}), (\ref{alphabeta1B}), one writes
\begin{equation}
	\alpha+\beta= 2{\mathrm{i}\alpha''},
	\label{alpha+beta1}
\end{equation} 
in such a way the refractive indices (\ref{iso10}) are written as 
\begin{align}
	n_{\pm}&=\sqrt{\mu {\epsilon} + \mu^{2} \left(\alpha''+\frac{\Sigma}{2\omega}\right)^{2}} \pm \mu  \left(\alpha''+\frac{\Sigma}{2\omega}\right). \label{isotropic-case-1-2-0-1}
\end{align}

\begin{figure}[h!]
	\begin{centering}
		\includegraphics[scale=0.68]{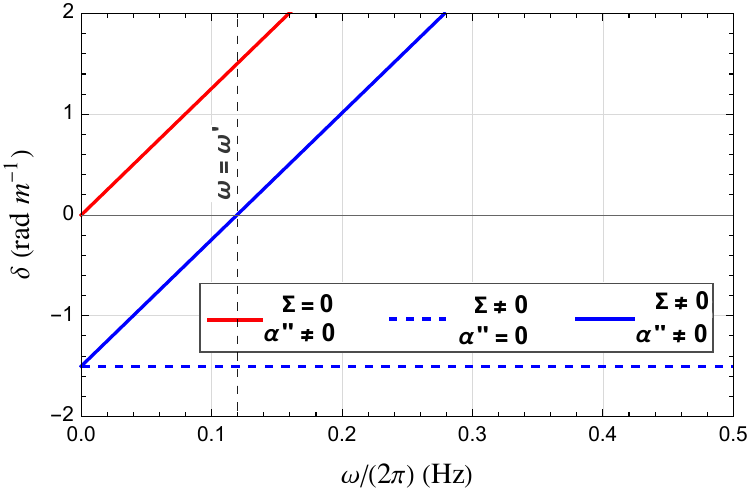}
		\par\end{centering}
	\caption{\label{plot-rotatory-dispersion-fullisotropic-case} Polarization rotation angle per length unit, as given in \eqref{isotropic-case-1-3}. Here, we have used $\mu=1$ $\mathrm{H}$ $\mathrm{m}^{-1}$, $\Sigma=3$ $\mathrm{\Omega}^{-1}$~$\mathrm{s}^{-1}$, and $|\alpha''|=2$ $\mathrm{F}$ $\mathrm{s}^{-1}$. The vertical dashed line is given by $\omega' /(2\pi) = 3/4$ $\mathrm{Hz}$, with $\omega'$ of \eqref{cutoff-frequency-full-isotropic-case}.}
\end{figure}

Replacing the latter indices in \eqref{rotatory-power-1}, one obtains the rotatory power 
	\begin{equation}
		\delta = - \frac{\mu  \Sigma}{2} -\mu \omega \alpha^{\prime \prime }, \label{isotropic-case-1-3}
\end{equation}
which receives a frequency-dependent contribution stemming from the chiral parameter $\alpha''$, which engenders a dispersive character. Regarding the possibility of a negative $\alpha ^{\prime \prime}$, the rotatory power is null at the frequency
\begin{align}
	\omega' &=  \frac{ \Sigma}{2|\alpha''|} , \label{cutoff-frequency-full-isotropic-case}
\end{align}
defining the value at which the RP sign reversion occurs. The RP (\ref{isotropic-case-1-3}) is depicted by the continuous blue line in Fig. \ref{plot-rotatory-dispersion-fullisotropic-case} (using SI units).

 The RP reversion here observed is not usual in ordinary, bi-isotropic or bi-anisotropic dielectrics, in media endowed with isotropic magnetic conductivity, as shown in \eqref{eq:rotatory-power}, or in conventional cold plasmas.  Such a reversion, however, is reported in graphene systems \cite{Poumirol}, Weyl metals and semimetals with low electron density with chiral conductivity \cite{Pesin,Dey-Nandy}, in rotating plasmas \cite{Gueroult,Gueroult2} and in axion chiral plasmas ruled by the MCFJ electrodynamics \cite{Filipe1,Filipe2}. The RP reversion, thus, constitutes a signature of chiral bi-isotropic media with magnetic current (or ruled by the timelike sector of the MCFJ electrodynamics). The examination of both the present and more intricate bi-anisotropic scenarios with magnetic current can be found in Ref. \cite{PedroPRB}.

\section{\label{section6}Final Remarks}

In this work, we have shown how to apply usual electromagnetic techniques to investigate optical properties of matter. Initially, in Sec.~\ref{section2}, we revisited some aspects of the classical framework to describe electromagnetic propagation in a usual dielectric with Ohmic conductivity. It was also applied to bi-isotropic and bi-anisotropic dielectrics, whose extended constitutive relations modify the dispersion equations and refractive indices. The corresponding propagating modes were determined, as well as optical effects (birefringence).

In Sec.~\ref{section3}, we have discussed the effects of the isotropic magnetic conductivity, ${\Sigma}$, on electromagnetic propagation in a dielectric substrate. In this case, the dispersion equation yields two distinct real indices $n_{\pm}$, associated with left- and right-handed circular polarization modes. These results indicate the absence of dissipation and are independent of the propagation direction, implying an ``isotropic birefringence'' \cite{Pedro1}. The medium can be described as optically active having a frequency-independent specific rotary power $\Delta=-\mu \Sigma/2$. The Hermitian effective permittivity explains this loss-free scenario. Additional aspects of energy propagation can be analyzed in terms of the group velocity \cite{Gerasik} and also energy velocity \cite{Gerasik,Brillouin, Loudon,Sherman,Davidovich,Ruppin}, which is not spoiled by divergence and superluminal values at low frequencies. For details, see Ref. \cite{PedroPRB2024A}.

The systems and results here discussed have illustrated the suitability of the classical electrodynamics formalism to successfully describe optical properties of dielectric media ruled by usual and modified constitutive relations, including the magnetic current one. Thus, this formalism constitutes a reliable technique to investigate optical effects in new materials (in the classical regime).

\subsection*{Acknowledgments}

The authors express their gratitude to FAPEMA, CNPq, and CAPES (Brazilian
research agencies) for invaluable financial support. In particular, M.M.F. is supported by CNPq/Produtividade 311220/2019-3, CNPq/Universal/422527/2021-1 and FAPEMA/APP 12151/22. P.D.S.S. is grateful to grant CNPq/PDJ 150584/23. Furthermore, we are indebted to CAPES/Finance Code 001 and FAPEMA/POS-GRAD-02575/21.


\begin{thebibliography}{99}

\bibitem{Jackson} J.D. Jackson, \textit{Classical Electrodynamics}, 3rd ed. (John Wiley \& Sons, New York, 1999).

\bibitem{Zangwill} A.~Zangwill, \textit{Modern Electrodynamics} (Cambridge University Press, New York, 2012).

\bibitem{Landau} L.D. Landau and E.M. Lifshitz, \textit{Electrodynamics
	of continuous media, Course of Theoretical Physics, Volume 8}, 2nd ed. (Pergamon Press, New York, 1984).

\bibitem{Fowles} G. R. Fowles, \textit{Introduction to modern optics}, 2nd
ed. (Dover Publications, INC., New York, 1975).

\bibitem{Bain}  A. K. Bain, \textit{Crystal optics: properties and applications} (Wiley-VCH Verlag GmbH \& Co. KGaA, Germany, 2019).

\bibitem{Kharzeev1} D.E.~Kharzeev, The chiral magnetic effect and anomaly-induced transport, \href{https://doi.org/10.1016/j.ppnp.2014.01.002}{Prog. Part. Nucl. Phys. {\bf 75}, 133 (2014)};
D.E.~Kharzeev, J.~Liao, S.A.~Voloshin, and G.~Wang, Chiral magnetic and vortical effects in high-energy nuclear collisions -- A status report, \href{https://doi.org/10.1016/j.ppnp.2016.01.001}{Prog. Part. Nucl. Phys. \textbf{88}, 1 (2016)};
D.~Kharzeev, K.~Landsteiner, A.~Schmitt and H.U.~Yee, \textit{Strongly Interacting Matter in Magnetic Fields},
Lect. Notes Phys. \textbf{871} (Springer-Verlag, Berlin $\cdot$ Heidelberg, 2013).

\bibitem{Fukushima} K.~Fukushima, D.E.~Kharzeev, and H.J.~Warringa, Chiral magnetic effect, \href{https://doi.org/10.1103/PhysRevD.78.074033}{Phys. Rev. D \textbf{78}, 074033 (2008)}.

\bibitem{Gabriele} G.~Inghirami, M.~Mace, Y.~Hirono, L.~Del Zanna, D.E.~Kharzeev, and M.~Bleicher, Magnetic fields in heavy ion collisions: flow and charge transport, \href{https://doi.org/10.1140/epjc/s10052-020-7847-4}{Eur. Phys. J. C \textbf{80}, 293 (2020)}.

\bibitem{Schober} J.~Schober, A.~Brandenburg and I.~Rogachevskii, Chiral fermion asymmetry in high-energy plasma simulations, \href{https://doi.org/10.1080/03091929.2019.1591393}{Geophys. Astrophys. Fluid Dynamics \textbf{114}, 106 (2020)}.

\bibitem{Vilenkin} A.~Vilenkin, Equilibrium parity-violating current in a magnetic field, \href{https://doi.org/10.1103/PhysRevD.22.3080}{Phys. Rev. D \textbf{22}, 3080 (1980)};
A.~Vilenkin and D.A.~Leahy, Parity nonconservation and the origin of cosmic magnetic fields, \href{https://doi.org/10.1086/159706}{Astrophys. J. \textbf{254}, 77 (1982)}.



\bibitem{Maxim} M.~Dvornikov and V.B.~Semikoz, Influence of the turbulent motion on the chiral magnetic effect in the early universe, \href{https://doi.org/10.1103/PhysRevD.95.043538}{Phys. Rev. D \textbf{95}, 043538 (2017)}.




\bibitem{Leite} G.~Sigl and N.~Leite, Chiral magnetic effect in protoneutron stars and magnetic field spectral evolution, \href{https://doi.org/10.1088/1475-7516/2016/01/025}{JCAP \textbf{01}, 025 (2016)}.



\bibitem{Dvornikov} M.~Dvornikov and V.B.~Semikoz, Magnetic field instability in a neutron star driven by the electroweak electron-nucleon interaction versus the chiral magnetic effect, \href{https://doi.org/10.1103/PhysRevD.91.061301}{Phys. Rev. D \textbf{91}, 061301(R) (2015)}.



\bibitem{Bubnov} A.F.~Bubnov, N.V.~Gubina, and V.Ch.~Zhukovsky, Vacuum current induced by an axial-vector condensate and electron anomalous magnetic moment in a magnetic field, \href{https://doi.org/10.1103/PhysRevD.96.016011}{Phys. Rev. D \textbf{96}, 016011 (2017)}.



\bibitem{Akamatsu} Y.~Akamatsu and N.~Yamamoto, Chiral Plasma Instabilities, \href{https://doi.org/10.1103/PhysRevLett.111.052002}{Phys. Rev. Lett. \textbf{111}, 052002 (2013)};
A.~Boyarsky, O.~Ruchayskiy, and M.~Shaposhnikov, Long-Range Magnetic Fields in the Ground State of the Standard Model Plasma, \href{https://doi.org/10.1103/PhysRevLett.109.111602}{Phys. Rev. Lett. \textbf{109}, 111602 (2012)}.





\bibitem{Maxim1} M.~Dvornikov and V.B.~Semikoz, Instability of magnetic fields in electroweak plasma driven by neutrino asymmetries, \href{https://doi.org/10.1088/1475-7516/2014/05/002}{JCAP \textbf{05}, 002 (2014)};
M.~Dvornikov, Chiral magnetic effect in the presence of an external axial-vector field, \href{https://doi.org/10.1103/PhysRevD.98.036016}{Phys. Rev. D \textbf{98}, 036016 (2018)}.


\bibitem{Li} Q. Li, D. E. Kharzeev, C. Zhang, Y. Huang, I. Pletikosi\'c, A. V. Fedorov, R. D. Zhong, J. A. Schneeloch, G. D. Gu, and T. Valla, Chiral magnetic effect in ZrTe5, \href{https://doi.org/10.1038/nphys3648}{Nat. Phys. {\bf{12}}, 550 (2016)}.

	\bibitem{Xiaochun-Huang}
	X. Huang, L. Zhao, Y. Long, P. Wang, D. Chen, Z. Yang, H. Liang, M. Xue, H. Weng, Z. Fang, X. Dai, and G. Chen, Observation of the chiral-anomaly-induced negative magnetoresistance in 3D Weyl semimetal TaAs, \href{https://doi.org/10.1103/PhysRevX.5.031023}{Phys. Rev. X {\bf{5}}, 031023 (2015)}.
	
	

	
	
	
	
	
	
\bibitem{Burkov} A.A.~Burkov, Chiral anomaly and transport in Weyl metals, \href{https://doi.org/10.1088/0953-8984/27/11/113201}{J. Phys. Condens. Matter \textbf{27}, 113201 (2015)}.

\bibitem{Barnes} E. Barnes, J. J. Heremans, and Djordje Minic, Electromagnetic Signatures of the Chiral Anomaly in Weyl Semimetals, \href{10.1103/PhysRevLett.117.217204}{Phys. Rev. Lett. \textbf{117}, 217204 (2016)}.










\bibitem{Chang} M.-C.~Chang and M.-F.~Yang, Chiral magnetic effect in a two-band lattice model of Weyl semimetal, \href{https://doi.org/10.1103/PhysRevB.91.115203}{Phys. Rev. B \textbf{91}, 115203 (2015)}.

\bibitem{Wurff} E.C.I~van der Wurff and H.T.C.~Stoof, Anisotropic chiral magnetic effect from tilted Weyl cones, \href{https://doi.org/10.1103/PhysRevB.96.121116}{Phys. Rev. B \textbf{96}, 121116(R) (2017)}.


\bibitem{Landsteiner} K.~Landsteiner, Anomalous transport of Weyl fermions in Weyl semimetals, \href{https://doi.org/10.1103/PhysRevB.89.075124}{Phys. Rev. B \textbf{89}, 075124 (2014)}.


\bibitem{Kaushik} S.~Kaushik and D.E.~Kharzeev, Quantum oscillations in the chiral magnetic conductivity, \href{https://doi.org/10.1103/PhysRevB.95.235136}{Phys. Rev. B \textbf{95}, 235136 (2017)}.



\bibitem{Ruiz} A.~Mart\'{i}n-Ruiz, M.~Cambiaso, and L.F.~Urrutia, Electromagnetic fields induced by an electric charge near a Weyl semimetal, \href{https://doi.org/10.1103/PhysRevB.99.155142}{Phys. Rev. B \textbf{99}, 155142 (2019)}.



\bibitem{Kharzeev2} D.E.~Kharzeev and H.J.~Warringa, Chiral magnetic conductivity, \href{https://doi.org/10.1103/PhysRevD.80.034028}{Phys. Rev. D \textbf{80}, 034028 (2009)}; D.E.~Kharzeev, Chiral magnetic superconductivity, \href{https://doi.org/10.1051/epjconf/201713701011}{EPJ Web Conf. \textbf{137}, 01011 (2017)}.

\bibitem{Qing-Dong} Q.-D. Jiang, H. Jiang, H. Liu, Q.-F. Sun, and X. C. Xie, Chiral wave-packet in Weyl semimetals, \href{https://doi.org/10.1103/PhysRevB.93.195165}{Phys. Rev. B {\bf{93}}, 195165 (2016)}.
\bibitem{Qing-Dong2}Q.-D. Jiang, H. Jiang, H. Liu, Q.-F. Sun, and X. C. Xie, Topological Imbert-Fedorov shift in Weyl semimetals, \href{https://doi.org/10.1103/PhysRevLett.115.156602}{Phys. Rev. Lett. {\bf{115}}, 156602 (2015)}.
	
	\bibitem{CFJ} S.M.~Carroll, G.B.~Field, and R.~Jackiw, Limits on a Lorentz- and parity-violating modification of electrodynamics,
	\href{https://doi.org/10.1103/PhysRevD.41.1231}{Phys. Rev. D \textbf{41}, 1231 (1990)}.
	
	\bibitem{Colladay} D.~Colladay and V.A.~Kosteleck\'{y}, \textit{CPT} violation and the standard model,
	\href{https://doi.org/10.1103/PhysRevD.55.6760}{Phys. Rev. D \textbf{55}, 6760 (1997)};
	D.~Colladay and V.A.~Kosteleck\'{y}, Lorentz-violating extension of the standard model,
	\href{https://doi.org/10.1103/PhysRevD.58.116002}{Phys. Rev. D \textbf{58}, 116002 (1998)};
	S.R.~Coleman and S.L.~Glashow, High-energy tests of Lorentz invariance,
	\href{https://doi.org/10.1103/PhysRevD.59.116008}{Phys. Rev. D \textbf{59}, 116008 (1999)}.
	

{\bibitem{KDeng} K. Deng, J. S. Van Dyke, D. Minic, J. J. Heremans, and E. Barnes, Exploring self-consistency of the equations of axion electrodynamics in Weyl semimetals, \href{https://doi.org/10.1103/PhysRevB.104.075202}{Phys. Rev. B \textbf{104}, 075202 (2021)}.}


{\bibitem{Wilczek} F. Wilczek, Two Applications of Axion Electrodynamics, \href{https://doi.org/10.1103/PhysRevLett.58.1799}{Phys. Rev. Lett. \textbf{58}, 1799 (1987)}.}


\bibitem{Sekine} A. Sekine and K. Nomura, Axion electrodynamics in
topological materials, \href{https://doi.org/10.1063/5.0038804}{J. Appl.
	Phys. \textbf{129}, 141101 (2021).}

\bibitem{Tobar} M. E. Tobar, B. T. McAllister, and M. Goryachev, Modified
axion electrodynamics as impressed electromagnetic sources through
oscillating background polarization and magnetization, \href{https://doi.org/10.1016/j.dark.2019.100339}%
{Phys. Dark Universe \textbf{26}, 100339 (2019)}.

\bibitem{Paixao}J. M. A. Paix\~ao, L. P. R. Ospedal, M. J. Neves, and J. A. Helay\"el-Neto, The axion-photon mixing in non-linear electrodynamic scenarios, \href{https://doi.org/10.1007/JHEP10(2022)160}{J. High Energ. Phys. 2022, 160 (2022)}.

\bibitem{Barredo}E. Barredo-Alamilla, Daniel A. Bobylev, and Maxim A. Gorlach, Axion electrodynamics without Witten effect in metamaterials, \href{https://doi.org/10.1103/PhysRevB.109.195136}{Phys. Rev. B \textbf{109}, 195136 (2024)}.


\bibitem{Qiu} Z.~Qiu, G.~Cao and X.-G.~Huang, Electrodynamics of chiral matter,
\href{https://doi.org/10.1103/PhysRevD.95.036002}{Phys. Rev. D \textbf{95}, 036002 (2017)}.


\bibitem{Ruiz1}A. Mart\'in-Ruiz and C. A. Escobar, Casimir effect between ponderable media as modeled by the standard model extension, \href{https://doi.org/10.1103/PhysRevD.94.076010}{Phys. Rev. D \textbf{94}, 076010 (2016)}; A. Mart\'in-Ruiz and C. A. Escobar, Local effects of the quantum vacuum in Lorentz-violating electrodynamics, \href{https://doi.org/10.1103/PhysRevD.95.036011}{Phys. Rev. D \textbf{95}, 036011 (2017)}. 

\bibitem{Gomez}A. Gomez, A. Martín-Ruiz, Luis F. Urrutia, Effective electromagnetic actions for Lorentz violating theories exhibiting the axial anomaly, \href{10.1016/j.physletb.2022.137043}{Phys. Lett. B \textbf{829}, 137043 (2022)}.

\bibitem{Ruiz2}A.G\'omez, R. M. von Dossow, A. Mart\'{\i}n-Ruiz, L. F. Urrutia, Lorentz invariance violation and the $CPT$-odd electromagnetic response of a tilted anisotropic Weyl semimetal, \href{https://link.aps.org/doi/10.1103/PhysRevD.109.065005}{Phys. Rev. D \textbf{109}, 065005 (2024)}.

\bibitem{Marco}A. V. Kostelecky, R. Lehnert, N. McGinnis, M. Schreck, B. Seradjeh, Lorentz violation in Dirac and Weyl semimetals, 
\href{https://doi.org/10.48550/arXiv.2112.14293}{Phys. Rev. Research {\bf{4}}, 023106 (2022)}.

\bibitem{Pedro2} P. D. S. Silva, L. L. Santos, M. M. Ferreira, Jr., and M. Schreck, Effects of CPT-odd terms of dimensions three and five on electromagnetic propagation in continuous matter, \href{https://doi.org/10.1103/PhysRevD.104.116023}{Phy. Rev. D \textbf{104}, 116023 (2021)}.





\bibitem{Barron2} L. D. Barron, \textit{Molecular Light Scattering and Optical Activity}, 2nd ed. (Cambridge University Press, New York, 2004).

\bibitem{Hecht} E. Hecht, \textit{Optics}, 4nd ed. (Addison Wesley, San
Francisco, 2002).

\bibitem{Wagniere} G. H. Wagniere, \textit{On Chirality and the Universal Asymmetry: Reflections on Image and Mirror Image}, (Wiley-Vch, Zurich) (2007).

\bibitem{TangPRL} Y. Tang and A. E. Cohen,  Optical Chirality and Its Interaction with Matter, \href{https://doi.org/10.1103/PhysRevLett.104.163901}{Phys. Rev. Lett. {\bf{104}}, 163901 (2010)}.

\bibitem{Sihvola1} A. H. Sihvola and I. V. Lindell, Bi-isotropic
constitutive relations, \href{https://doi.org/10.1002/mop.4650040805}{%
	Microw. Opt. Technol. Lett., \textbf{4} (8), 295-297 (1991)}; 
\bibitem{Sihvola2} A. H. Sihvola and I. V. Lindell, Properties of
bi-isotropic Fresnel reflection coefficients, \href{https://www.sciencedirect.com/science/article/abs/pii/003040189290237L?via\%3Dihub}{Optics Communications 89, 11992)}; S. Ougier, I. Chenerie, A. Sihvola, and A. Priou,
Propagation in bi-isotropic media: effect of different formalisms on the
propagation analysis, \href{http://www.jpier.org/PIER/pier.php?paper=9301010}%
{Progress In Electromagnetics Research \textbf{09}, 19 (1994)}.


\bibitem{Sihvola3} S. Ougier, I. Chenerie, A. Sihvola, and A. Priou,
Propagation in bi-isotropic media: effect of different formalisms on the
propagation analysis, \href{http://www.jpier.org/PIER/pier.php?paper=9301010}%
{Progress In Electromagnetics Research \textbf{09}, 19 (1994)}.

\bibitem{Sihvola4} I. V. Lindell, A. H. Sihvola, S. A. Tretyakov, and A. J. Viitanen,
\textit{Electromagnetic Waves in Chiral and Bi-Isotropic Media} (Artech House, Boston, 1993).



\bibitem{Bianiso} P. Hillion, Manifestly covariant formalism for
electromagnetism in chiral media, \href{https://doi.org/10.1103/PhysRevE.47.1365}%
{Phys. Rev. E \textbf{47}, 1365 (1993)};  I. Yakov, Dispersion relation for electromagnetic waves in anisotropic media, \href{https://doi.org/10.1016/j.physleta.2009.12.071}{Phys. Lett. A \textbf{374}, 1113 (2010)}; N.J. Damaskos, A.L. Maffett and P.L.E. Uslenghi, Dispersion relation for general anisotropic media, \href{https://ieeexplore.ieee.org/document/1142905}{IEEE Trans. Antennas Propagat. AP-30, 991 (1982)}.


\bibitem{Kong} J. A. Kong, \textit{Electromagnetic Wave Theory} (Wiley, New York, 1986).

\bibitem{Aladadi} Y. T. Aladadi and M. A. S. Alkanhal, Classification and
characterization of electromagnetic materials, \href{https://doi.org/10.1038/s41598-020-68298-3}%
{Sci. Rep. \textbf{10}, 11406 (2020)}.

\bibitem{Mahmood} W. Mahmood and Q. Zhao, The Double Jones Birefringence in Magneto-electric Medium, \href{https://doi.org/10.1038/srep13963} {Sci. Rep.  \textbf{5}, 13963 (2015)}.


\bibitem{Pedro3}
P. D. S.~Silva, R. Casana, and M. M.~Ferreira Jr., Symmetric and antisymmetric constitutive tensors for bi-isotropic and bi-anisotropic media, \href{https://10.1103/PhysRevA.106.042205}{Phys. Rev. A {\bf{106}},  042205 (2022)}.


\bibitem{Bennett}H. S. Bennett, E. A. Stern, Faraday effect in solids. \href{https://doi.org/10.1103/PhysRev.137.A448}{Phys. Rev. \textbf{137}, A448--A461 (1965)}; L. M. Roth. Theory of the Faraday effect in solids, \href{https://doi.org/10.1103/PhysRev.133.A542}{Phys. Rev. \textbf{133}, A542--A553 (1964)}. \bibitem{Porter}W. S. Porter and E. M. Bock Jr., Faraday effect in a plasma, \href{https://doi.org/10.1119/1.1971154}{Am. J. Phys. \textbf{33}, 1070 (1965)}.
\bibitem{Shibata} J. Shibata, A. Takeuchi, H. Kohno, and G. Tatara, Theory of electromagnetic wave propagation in ferromagnetic Rashba conductor, \href{ https://doi.org/10.1063/1.5011130}{J. App. Phy. \textbf{123}, 063902 (2018)}.	
\bibitem{Condon} E. U. Condon, Theories of Optical Rotatory Power, \href{https://doi.org/10.1103/RevModPhys.9.432}{Rev. Mod. Phys. {\bf{9}}, 432 (1937)}.


\bibitem{Rado} G. T. Rado and V. J. Folen, Observation of the Magnetically Induced Magnetoelectric Effect and Evidence for Antiferromagnetic Domains, \href{https://doi.org/10.1103/PhysRevLett.7.310}{ Phys. Rev. Lett. \textbf{7}, 310 (1961)}; D. N. Astrov, Magnetoelectric effect in chromium oxide, \href{http://jetp.ras.ru/cgi-bin/e/index/e/13/4/p729?a=list}{Soviet Physics JETP 13, 729 (1961)}.

\bibitem{Jelinek}
L. Jelinek, R. Marqu\'es, F. Mesa, and J. D. Baena, Periodic arrangements of chiral scatterers providing negative refractive index bi-isotropic media, \href{https://doi.org/10.1103/PhysRevB.77.205110}{Phys. Rev. B 77, 205110 (2008)}.




\bibitem{Zou} Zheng-Wei Zuo, Dong-Bo Ling, L.Sheng, D.Y.Xing, Optical properties for topological insulators with metamaterials, \href{http://dx.doi.org/10.1016/j.physleta.2013.09.004}{Phys. Lett. A \textbf{377}, 2909 (2013).}









\bibitem{Urrutia} A. Mart\'in-Ruiz, M. Cambiaso, and L. F. Urrutia, The
magnetoelectric coupling in Electrodynamics. \href{https://www.worldscientific.com/doi/abs/10.1142/S0217751X19410021}%
{ Int. J. Mod. Phys. A \textbf{34}, 1941002 (2019)}; A.~Mart\'{\i}n-Ruiz, M.~Cambiaso, and L.F.~Urrutia, Electro- and magnetostatics of topological insulators as modeled by planar,
spherical, and cylindrical $\theta$ boundaries: Green's function
approach, \href{https://journals.aps.org/prd/abstract/10.1103/PhysRevD.93.045022}
{Phys. Rev. D \textbf{93}, 045022 (2016)}.

\bibitem{Lakhtakia} A. Lakhtakia and T. G. Mackay, Classical electromagnetic
model of surface states in topological insulators, \href{https://doi.org/10.1117/1.JNP.10.033004}%
{J. Nanophoton. \textbf{10} (3), 033004 (2016)}.

\bibitem{Winder} T. M. Melo, D. R. Viana, W. A. Moura-Melo, J. M. Fonseca,
A. R. Pereira, Topological cutoff frequency in a slab waveguide: Penetration
length in topological insulator walls, \href{https://doi.org/10.1016/j.physleta.2015.12.041}%
{Phys, Lett. A \textbf{380}, 973 (2016)}.








\bibitem{Li1} R. Li, J. Wang, Xiao-Liang Qi and S.-C. Zhang, Dynamical axion
field in topological magnetic insulators, \href{https://doi.org/10.1038/NPHYS1534}
{Nature Phys. \textbf{6}, 284 (2010)}.






\bibitem{Tokura} Y.Tokura, K.Yasuda, A. Tsukazaki,  Magnetic topological insulators. \href{https://doi.org/10.1038/s42254-018-0011-5}{Nat. Rev. Phys. \textbf{1}, 126 (2019)}.


\bibitem{Li-Cao} Z.-X. Li, Yunshan Cao, Peng Yan, Topological insulators and
semimetals in classical magnetic systems, \href{https://doi.org/10.1016/j.physrep.2021.02.003}%
{Phys. Report \textbf{915}, 1 (2021)}.



\bibitem{BorgesAxion} L. H. C. Borges, A. G. Dias, A. F. Ferrari, J. R.
Nascimento, A. Yu. Petrov, Generation of Axion-Like Couplings via Quantum
Corrections in a Lorentz Violating Background, \href{https://doi.org/10.1103/PhysRevD.89.045005}%
{ Phys. Rev. D \textbf{89}, 045005 (2014)}.


\bibitem{Silveirinha} F. R. Prud\^encio and M. G. Silveirinha, Optical isolation of circularly polarized light with a spontaneous magnetoelectric effect, \href{https://doi.org/10.1103/PhysRevA.93.043846} {Phys. Rev. A {\bf{93}}, 043846 (2016)}.


\bibitem{Casimir} R. Zhao, J. Zhou, Th. Koschny, E. N. Economou, and C. M. Soukoulis, Repulsive Casimir Force in Chiral Metamaterials, \href{https:// 10.1103/PhysRevLett.103.103602} {Phys. Rev. Lett. \textbf{103}, 103602 (2009)}; M. G. Silveirinha and S. I. Maslovski, Comment on Repulsive Casimir Force in Chiral Metamaterials, \href{https:// https://doi.org/10.1103/PhysRevLett.105.189301} {Phys. Rev. Lett. \textbf{105}, 189301 (2010).}

\bibitem{Casimir2} T. Schoger and G.-L. Ingold, Switching the sign of the Casimir force between two perfect electromagnetic conductor spheres, \href{https://doi.org/10.1103/PhysRevA.109.052815}{Phys. Rev. A \textbf{109}, 052815 (2024)}.

\bibitem{Casimir3}R. Zhao, J. Zhou, Th. Koschny, E. N. Economou, and C. M. Soukoulis, Repulsive Casimir Force in Chiral Metamaterials, \href{https:// 10.1103/PhysRevLett.103.103602} {Phys. Rev. Lett. \textbf{103}, 103602 (2009)}; Z. Dai and Q.-D. Jiang, A universal roadmap for searching repulsive Casimir forces between magneto-electric materials, arXiv: 2403.00740.



\bibitem{Darinskii}A. N. Darinskii, Surface plasmon polaritons in metal films on anisotropic and bianisotropic substrates, \href{https://doi.org/10.1103/PhysRevA.104.023507}{Phys. Rev. A \textbf{104}, 023507 (2021)}.





\bibitem{Chang1} Ming-Che Chang and Min-Fong Yang, Optical signature of
topological insulators, \href{https:10.1103/PhysRevB.80.113304}{Phys. Rev. B \textbf{80}, 113304 (2009)}; L. Ohnoutek \textit{et. al.}, Strong interband Faraday rotation in 3D topological insulator $\mathrm{Bi}_{2}\mathrm{Se}_{3}$, \href{https://doi.org/10.1038/srep19087} {Sci. Rep.\textbf{6}, 19087 (2016).}





\bibitem{Tse}W.-K. Tse and A. H. MacDonald, Giant magneto-optical Kerr effect and universal Faraday effect in thin-film topological insulators, \href{https://doi.org/10.1103/PhysRevLett.105.057401}{Phys. Rev. Lett. {\bf{105}}, 057401 (2010)}; Magneto-optical and magnetoelectric effects of topological insulators in quantizing magnetic fields,	 \href{https://doi.org/10.1103/PhysRevB.82.161104} {Phys. Rev. B {\bf{82}}, 161104 (2010)};  Magneto-optical Faraday and Kerr effects in topological insulator films and in other layered quantized Hall systems, \href{https://doi.org/10.1103/PhysRevB.84.205327}{Phys. Rev. B {\bf{84}}, 205327 (2011)}.


\bibitem{Crasee}I. Crassee, J. Levallois, A. L. Walter, M. Ostler, A. Bostwick, E. Rotenberg, T. Seyller, D. van der Marel, and A. B. Kuzmenko, Giant Faraday rotation in single- and multilayer graphene, \href{https://doi.org/10.1038/nphys1816}{Nature Phys. {\bf{7}}, 48 (2011)}; R. Shimano, G. Yumoto, J. Y. Yoo, R. Matsunaga, S. Tanabe, H. Hibino, T. Morimoto and H. Aoki, Quantum Faraday and Kerr rotations in graphene, \href{https://doi.org/10.1038/ncomms2866}{Nature Comm. {\bf{4}}, 1841 (2013)}.





\bibitem{Ohnoutek}
L. Ohnoutek et al., Strong interband Faraday rotation in 3D topological insulator $Bi_{2}Se_{3}$, \href{https://doi.org/10.1038/srep19087}{Sci. Rep. 6, 19087 (2016)}.




\bibitem{Liang}
L. Wu, M. Salehi, N. Koirala, J. Moon, S. Oh, and N. P. Armitage, Quantized Faraday and Kerr rotation and axion electrodynamics of a 3D topological insulator, \href{https://doi.org/10.1126/science.aaf5541}{Science 354, 6316 (2016)}.



\bibitem{Kamenetskii} E. O. Kamenetskii, Energy balance
equation for electromagnetic waves in bianisotropic media,
\href{https://doi.org/10.1103/PhysRevE.54.4359} {Phys. Rev. E \textbf{54}, 4359 (1996)}.




\bibitem{Carvalho} C. A. A. de Carvalho, The relativistic electron gas: a
candidate for nature's left-handed material, \href{https://doi.org/10.1103/PhysRevD.93.105005}%
{ Phys. Rev. D \textbf{93}, 105005 (2016)}; E. Reyes-G\'omez, L. E. Oliveira
and C. A. A. de Carvalho, The electromagnetic response of a relativistic
Fermi gas at finite temperatures: Applications to condensed-matter systems,
\href{https://doi.org/10.1209/0295-5075/114/17009}{ EPL \textbf{114}, 17009
	(2016)}.

\bibitem{Lin} R.-Y. Zhang, Y.-W. Zhai, S.-R. Lin, Q. Zhao, W. Wen, M.-L. Ge, Time Circular Birefringence in
Time-Dependent Magnetoelectric Media,
\href{http://10.1038/srep13673 (2015)} {Sci. Rep. \textbf{5}, 13673 (2015)}; S.-R. Lin, R.-Y. Zhang, Y.-R. Ma, W. Jia, Q. Zhao, Electromagnetic wave propagation in time-dependent media with antisymmetric magnetoelectric coupling, \href{http://dx.doi.org/10.1016/j.physleta.2016.05.050} {Phys. Lett. A \textbf{380}, 2582 (2016)}.

\bibitem{Halterman} K. Halterman, M. Alidoust and A. Zyuzin,
Epsilon-near-zero response and tunable perfect absorption in Weyl
semimetals, \href{https://doi.org/10.1103/PhysRevB.98.085109}{Phys. Rev. B
	\textbf{98}, 085109 (2018)}.

\bibitem{Zu} R. Zu, M. Gu, L. Min, C. Hu, N. Ni, Z. Mao, J. M. Rondinelli
and V. Gopalan, Comprehensive anisotropic linear optical properties of Weyl
semimetals TaAs and NbAs, \href{https://doi.org/10.1103/PhysRevB.103.165137}{Phys. Rev. B {\bf{103}}, 165137 (2021)}.

\bibitem{Krupka1} J. Krupka, Measurement of the complex permittivity,
initial permeability, permeability tensor and ferromagnetic linewidth of
gyromagnetic materials, \href{https://doi.org/10.1088/1361-6501/aacf5d}{%
	Meas. Sci. Technol. \textbf{29}, 092001 (2018)}.

\bibitem{Krupka2} J. Krupka, A. Pacewicz, B. Salski, P. Kopyt, J. Bourhill,
M. Goryachev and M. Tobar, Electrodynamic improvements to the theory of
magnetostatic modes in ferrimagnetic spheres and their applications to
saturation magnetization measurements, \href{https://doi.org/10.1016/j.jmmm.2019.165331}%
{J. Magnetism and Magnetic Materials, \textbf{487}, 165331 (2019)}.

\bibitem{Hillion} P. Hillion, Manifestly covariant formalism for
electromagnetism in chiral media, \href{https://doi.org/10.1103/PhysRevE.47.1365}%
{Phys. Rev. E \textbf{47}, 1365 (1993)}.

\bibitem{Yakov} 
Y. Itin, Dispersion relation for electromagnetic waves in
anisotropic media, \href{https://doi.org/10.1016/j.physleta.2009.12.071}{%
	Phys. Lett. A \textbf{374}, 1113 (2010)}.




\bibitem{Damaskos} N.J. Damaskos, A.L. Maffett and P.L.E. Uslenghi,
Dispersion relation for general anisotropic media, \href{https://ieeexplore.ieee.org/document/1142905}{IEEE Trans. Antennas Propagat. AP-30, 991 (1982)}.








\bibitem{Kurumaji} T. Kurumaji, Y. Takahashi, J. Fujioka, R. Masuda, H. Shishikura, S. Ishiwata, and Y. Tokura, Optical Magnetoelectric Resonance in a Polar Magnet $(Fe,Zn)_{2}2Mo_{3}O_{8}$ with Axion-Type Coupling, \href{https://doi.org/10.1103/PhysRevLett.119.077206}{Phys. Rev. Lett. \textbf{119}, 077206 (2017).}





\bibitem{Takahashi} Y. Takahashi, R. Shimano, Y. Kaneko, H. Murakawa and Y. Tokura, Magnetoelectric resonance with electromagnons in a perovskite helimagnet,  \href{https://doi.org/10.1038/NPHYS2161}{Nature Phys. \textbf{8}, 121 (2012)}; S. Iguchi, R. Masuda, S. Seki, Y. Tokura, Y. Takahashi, Enhanced gyrotropic birefringence and natural optical activity on electromagnon resonance in a helimagnet, \href{https://doi.org/10.1038/s41467-021-26953-x}{Nature Communications \textbf{12}, 6674 (2021)}.




\bibitem{Pedro1} P. D. S. Silva, M. M. Ferreira Jr., M. Schreck, and L. F. Urrutia,  Magnetic-conductivity effects on electromagnetic propagation in
	dispersive matter, \href{https://journals.aps.org/prd/abstract/10.1103/PhysRevD.102.076001}{Phys. Rev. D {\bf{102}}, 076001 (2020)}.


\bibitem{PedroPRB2024A} P. D. S. Silva, M.J. Neves, M. M. Ferreira Jr.,  Optical properties and energy propagation in a dielectric medium supporting magnetic current, \href{https://doi.org/10.1103/PhysRevB.109.184439}{Phys. Rev. B {\bf{109}}, 184439 (2024)}.

\bibitem{PedroPRB2024B} P. D. S. Silva, M.J. Neves, M. M. Ferreira Jr.,  Drude-Lorentz dielectric in the presence of a magnetic current density., \href{https://doi.org/10.1103/PhysRevB.109.184444}{Phys. Rev. B {\bf{109}}, 184444 (2024)}.

\bibitem{PedroPRB}
P. D. S.Silva and M. M.~Ferreira Jr., Rotatory power reversal induced by magnetic current in bi-isotropic media, \href{https://10.1103/PhysRevB.106.144430} {Phys. Rev. B {\bf{106}}, 144430 (2022)};  {Erratum: Rotatory power reversal induced by magnetic current in bi-isotropic media [Phys. Rev. B 106, 144430 (2022)]}, \href{https://doi.org/10.1103/PhysRevB.107.179902}{Phys. Rev. B {\bf{107}}, 179902 (2023)}.

\bibitem{Kaushik2}
S.~Kaushik, D.E.~Kharzeev, and E.J.~Philip, Transverse chiral magnetic photocurrent induced by linearly polarized light in symmetric Weyl semimetals, \href{https://doi.org/10.1103/PhysRevResearch.2.042011}{Phys. Rev. Research {\bf{2}}, 042011(R) (2020)}.


\bibitem {Poumirol} J.-M. Poumirol, P. Q. Liu, T. M. Slipchenko, A. Y. Nikitin, L. Martin-Morento, J. Faist, and A. B. Kuzmenko, Electrically controlled terahertz magneto-optical phenomena in continuous and patterned graphene, \href{https://doi.org/10.1038/ncomms14626}{Nat Commun 8, 14626 (2017)}.
\bibitem{Pesin} J. Ma, and D. A. Pesin, Dynamic chiral magnetic effect and Faraday rotation in macroscopically disordered helical metals, \href{https://doi.org/10.1103/PhysRevLett.118.107401}{Phys. Rev. Lett. \textbf{118}, 107401 (2017)}.

\bibitem{Dey-Nandy} U. Dey, S. Nandy, and A. Taraphder, Dynamic chiral magnetic effect and anisotropic natural optical activity of tilted Weyl semimetals, \href{https://doi.org/10.1038/s41598-020-59385-6}{Sci Rep 10, 2699 (2020)}.

\bibitem {Gueroult} R. Gueroult, J.-M. Rax, and N. J. Fisch, Enhanced tuneable rotatory power in a rotating plasma, \href{https://doi.org/10.1103/PhysRevE.102.051202}{Phys. Rev. E 102, 051202(R) (2020)}.

\bibitem{Gueroult2} R. Gueroult, Y. Shi, J-M. Rax, and N. J. Fisch, Determining the rotation direction in pulsars, \href{https://doi.org/10.1038/s41467-019-11243-4}{Nat. Commun. {\bf{10}}, 3232 (2019)}.


\bibitem{Filipe1} Filipe S. Ribeiro, Pedro D.S. Silva, M.M.Ferreira Jr., Cold plasma modes in the chiral Maxwell-Carroll-Field-Jackiw electrodynamics, \href{https://doi.org/10.1103/PhysRevD.107.096018}{Phys. Rev. D{ \bf{107}}, 096018 (2023)}.

\bibitem{Filipe2} Filipe S. Ribeiro, Pedro D.S. Silva, M.M.Ferreira Jr., Anisotropic cold plasma modes in chiral vector Maxwell-Carroll-Field-Jackiw electrodynamics, \href{https://doi.org/10.1103/PhysRevD.109.076003}{Phys. Rev. D{ \bf{109}}, 076003 (2024)}. 



\bibitem{Gerasik} V. Gerasik and M. Stastna, Complex group velocity and energy transport in absorbing media, \href{https://doi.org/10.1103/PhysRevE.81.056602}{Phys. Rev. E {\bf{81}}, 056602 (2010)}.





\bibitem{Brillouin}  L. Brillouin, Wave Propagation and Group Velocity (Academic Press, New York, 1960).
	
	
	
	\bibitem{Loudon} R. Loudon, The propagation of electromagnetic energy through an absorbing dielectric, \href{https://doi.org/10.1088/0305-4470/3/3/008}{J. Phys. A: Gen. Phys.{\bf{3}}, 233 (1970)}.
	
	\bibitem{Sherman}
	G. C. Sherman and K. E. Oughstun, Energy-velocity description of pulse propagation in absorbing, dispersive dielectrics, \href{https://doi.org/10.1364/JOSAB.12.000229}{J. Opt. Soc. Am. B \textbf{12}, 229-247 (1995)}.
	
	\bibitem{Davidovich} M.V. Davidovich, Electromagnetic energy density and velocity in a medium with anomalous positive dispersion, \href{https://doi.org/10.1134/S106378500611023X}{Tech. Phys. Lett. \textbf{32}, 982–986 (2006).}
	

	
	\bibitem{Ruppin}
	R. Ruppin, Electromagnetic energy density in a dispersive and absorptive material, \href{https://doi.org/10.1016/S0375-9601(01)00838-6}{Phys. Lett. A {\bf{299}}, 309-312 (2002)}.













\end{thebibliography}
\end{document}